\documentclass[12pt,preprint]{aastex}
\usepackage{epsfig}
\def\vkm{km s$^{-1}$}

\def\mJyb{mJy Beam$^{-1}$}
\def\Jybk{Jy beam$^{-1}$ km s$^{-1}$}
\def\cmc{cm$^{-3}$}

\def\cms{cm$^{-2}$}

\def\arcs#1{$#1''$}
\def\arcsa#1#2{$#1^{\prime\prime}_{^\textrm{.}}#2$}
\def\arcsaq#1#2{#1^{\prime\prime}_{^\textrm{.}}#2}

\def\leftblank#1{}

\def\nHC3N{HC$_3$N}
\def\H13CCCN{H$^{13}$CCCN}
\def\HC13CCN{HC$^{13}$CCN}
\def\HCC13CN{HCC$^{13}$CN}

\def\CH2CN{CH$_2$CN}
\def\C2H3CN{CH$_2$CHCN}

\def\vn{$v=0$}
\def\vn{}
\def\v#1#2{$v_{#1}=#2$}
\def\vexp{v_\textrm{\scriptsize exp}}
\def\lpeak{T_{B}^p}
\newcommand{\degree}{$^{\circ}$}
\def\cms{\textrm {cm}$^{-2}$}
\def\cmc{\textrm {cm}$^{-3}$}
\def\ro{r_0}

\def\mH2{m_{\textrm{\scriptsize H}_2}}
\def\nH2{n_{\textrm{\scriptsize H}_2}}
\def\vkm{km s$^{-1}$}
\def\smpy{M$_\odot$ yr$^{-1}$}

\def\solarM{M_\odot}
\def\solarL{L_\odot}

\def\putfig#1#2#3{\epsfig{scale=#1,angle=#2,figure=#3}}
\begin{document}

\title{Mapping the Central Region of the PPN CRL 618 at Sub-arcsecond
Resolution at 350 GHz}

\author{Chin-Fei Lee\altaffilmark{1}, Chun-Hui Yang\altaffilmark{1},
Raghvendra Sahai\altaffilmark{2}, and Carmen S\'anchez
Contreras\altaffilmark{3}}
\altaffiltext{1}{Academia Sinica Institute of Astronomy and Astrophysics,
P.O.  Box 23-141, Taipei 106, Taiwan}
\altaffiltext{2}{
Jet Propulsion Laboratory, MS 183-900, California Institute of Technology,
Pasadena, CA 91109, USA}
\altaffiltext{3}{
Astrobiology Center (CSIC-INTA), ESAC Campus, E-28691 Villanueva de la
Canada, Madrid, Spain
}

\begin{abstract}

CRL 618 is a well-studied pre-planetary nebula.  We have mapped its central
region in continuum and molecular lines with the Submillimeter Array at 350
GHz at $\sim$ \arcsa{0}{3} to \arcsa{0}{5} resolutions.  Two components are
seen in 350 GHz continuum: (1) a compact emission at the center tracing the
dense inner part of the H II region previously detected in 23 GHz continuum
and it may trace a fast ionized wind at the base, and (2) an extended
thermal dust emission surrounding the H II region, tracing the dense core
previously detected in \nHC3N at the center of the circumstellar envelope. 
The dense core is dusty and may contain mm-sized dust grains.  It may have a
density enhancement in the equatorial plane.  It is also detected in carbon
chain molecules \nHC3N and HCN, and their isotopologues, with higher
excitation lines tracing closer to the central star.  It is also detected in
\C2H3CN{} toward the innermost part.  Most of the emission detected here
arises within $\sim$ 630 AU (\arcsa{0}{7}) from the central star.  A simple
radiative transfer model is used to derive the kinematics, physical
conditions, and the chemical abundances in the dense core.  The dense core
is expanding and accelerating, with the velocity increasing roughly linearly
from $\sim$ 3 \vkm{} in the innermost part to $\sim$ 16 \vkm{} at 630 AU. 
The mass-loss rate in the dense core is extremely high with a value of $\sim
1.15\times 10^{-3}$ \smpy{}.  The dense core has a mass of $\sim$ 0.47
$\solarM{}$ and a dynamical age of $\sim$ 400 yrs.  It could result from a
recent enhanced heavy mass-loss episode that ends the AGB phase.  The
isotopic ratios of $^{12}$C/$^{13}$C and $^{14}$N/$^{15}$N are 9$\pm4$ and
150$\pm50$, respectively, both lower than the solar values.

\end{abstract}


\section{Introduction}

Most stars are low- and intermediate-mass stars and they end their lives the
same way as the Sun.  They first evolves into red giant branch (RGB) stars
and then asymptotic giant branch (AGB) stars with intense mass loss,
producing atomic and molecular circumstellar envelopes around them. 
Eventually, they evolves into white dwarfs hot enough to photoionize the
envelopes, forming spectacular emission nebulae called planetary nebulae
(PNe).  PNe are mostly bipolar and multipolar, but their shaping mechanism
is still uncertain.  Pre-planetary nebulae (PPNe) are transient objects in
the transition phase between the AGB phase and PN phase.  Their central
stars are post-AGB stars, which will evolve rapidly into the hot white
dwarfs, turning the PPNe into PNe in less than 1000 yrs.  PPNe are also
mostly bipolar and multipolar \citep{Sahai2007}, indicating that the shaping
of PNe must have started earlier in the PPN phase, see, e.g, the simulations
in \citet{Lee2003} and \citet{Lee2009}.  Since the circumstellar envelopes
may help shaping the structures of the PPNe \citep{Balick2002}, it is
important to determine their physical and kinematic properties.

CRL 618 is a nearby ($\sim$ 900 pc) well-studied PPN, with the morphological
classification Mcw,ml,h(e,a) based on HST imaging \citep{Sahai2007}, where
M=primary class is multipolar; c=outflow lobes are closed at their ends,
w=obscuring waist; ml=minor outflow lobes are present; h(e,a)=elongated halo
is present around nebula and shows (some) circular arc structures.  The
radio image in 23 GHz continuum showed a compact H II region close to the central
star \citep{Kwok1984,Martin1993}, suggesting that this PPN has started to evolve into
a PN at the center.  The optical image showed two pairs of collimated
outflow lobes in the east-west direction, expanding rapidly away from the
star \citep{Trammell2002,Sanchez2002}.  Since CRL 618 is a C-rich star,
single-dish molecular line surveys detected many lines from carbon chain
molecules HCN, \nHC3N, and their isotopologues at various vibrational
states, arising from a circumstellar envelope that expands at $\sim$ 5-18
\vkm{} \citep{Wyrowski2003,Pardo2004,Pardo2007b}. 
A total of 1736 lines of \nHC3N, its
isotopologues, and its vibrationally excited states have been previously
reported by \citet{Pardo2004,Pardo2007b} and \citet{Pardo2007a}.
This source is also the first one in which benzene and polyacetylenes
were detected in space \citep{Cernicharo2001a,Cernicharo2001b,Fonfria2011}.
The envelope was found to
have a $^{12}$C/$^{13}$C ratio of 10-15 \citep{Wyrowski2003,Pardo2004}, much
lower than the solar value.

In the interferometric observations in HCO$^{+}$ J=1-0 at $\sim$ \arcs{2}
resolution, \citet{SanchezS2004} found in the envelope a large expanding
torus with a diameter of $\sim$ \arcs{11} (10,000 AU) perpendicular to the
outflow axis.  In the interferometric observations in CO J=2-1 and \nHC3N{}
J=25-24 at $\sim$ \arcs{1} resolution, the envelope was resolved into an
extended round halo and a compact dense torus-like core near the central
star aligned with the large expanding torus (S{\'a}nchez Contreras et al. 
2004, hereafter \citet{Sanchez2004}).  As argued by \citet{Sanchez2004}, the
dense core may result from a recent heavy mass loss from the central star
and it may help shaping the PPN.  In order to check these possibilities, we
present our observations of the central region at $\sim$ 2-3 times higher
resolution, obtained with the Submillimeter Array (SMA) in the 350 GHz band. 
In our observations, many lines are also detected in the dense core, arising
from the carbon chain molecules, allowing us to refine not only the physical
and kinematic properties, but also the chemical properties of the dense core
at higher resolution.  In particular, our observations provide a much
higher angular resolution than that of \citet{Pardo2004,Pardo2007b},
allowing us to directly distinguish the different contributions to the
molecular emission, study the detailed spatial distribution of \nHC3N{}, and
thus to understand the chemical processes at work \citep{Cernicharo2004}. 
Moreover, the dense core and the H II region can also be seen and studied in 350 GHz
continuum.

\section{Observations}

%The phase center of
%the map was taken at R.A.=\ra{04}{42}{53}{64}, decl.=\dec{36}{06}{53}{4}. 

The observations toward CRL 618 were carried out on 2011 January 23 and
February 4 with the SMA in the very extended and extended configurations,
respectively.  The detailed information of the SMA can be found in 
\cite{Ho2004}.  In these observations, the receivers were setup to have the
following two frequency ranges: 342.104--346.065 in the lower sideband and
354.115--358.068 GHz in the upper sideband.  These frequency ranges covered
the lines of CO, CS, HCO$^+$, \nHC3N and HCN and their isotopologues,
simultaneously with the 350 GHz continuum.  The correlator was setup to have
a velocity resolution from 0.35 to 1.41 km s$^{-1}$ per channel.  One single
pointing was used to observe the central region of CRL 618, with a field of
view of \arcs{34}.  Six and seven antennas were used in the very extended
and extended configurations, respectively.  The baseline length, after
combining the two configurations, ranged from 45 to 460 m.  The observations
were interleaved every 5 minutes with nearby gain calibrators, 3C 84 and 3C
111, to track the phase variations over time.  However, only 3C 111 was used
for the gain calibration because it is much closer to the source and was
already bright enough.  The bandpass calibrator was the quasar 3C 279, and
the flux calibrator was Titan.  The total on-source time was $\sim$ 5 hrs in
each configuration.  The system temperature ranges were from 220 to 660 K
and from 250 to 900 K in the very extended and extended configurations,
respectively.

The visibility data were calibrated with the MIR package.  The flux
uncertainty was estimated to be $\sim$ 20\%.  The continuum band was
obtained from the line-free channels.  The calibrated visibility data were
imaged with the MIRIAD package.  The dirty maps that were produced from the
calibrated visibility data were CLEANed using the Steer clean method,
producing the CLEAN component maps.  The final maps were obtained by
restoring the CLEAN component maps with a synthesized (Gaussian) beam fitted
to the main lobe of the dirty beam.  With natural weighting, the synthesized
beam has a size of \arcsa{0}{53}$\times$\arcsa{0}{36} at a position angle
(P.A.) of $\sim$ 83\degree{}.  The rms noise levels are $\sim$ 60 \mJyb{}
for the channel maps with a velocity resolution of 1.4 \vkm{}, and 3.7
\mJyb{} for the continuum map.  The velocities of the channel maps are LSR.

\section{Observational Results} 

\subsection{Continuum: Dense core and H II region} \label{sec:cont}

%from conth.mem  "(0.1,-0.09,-0.1,0.1)"

\begin{figure} [!hbp]
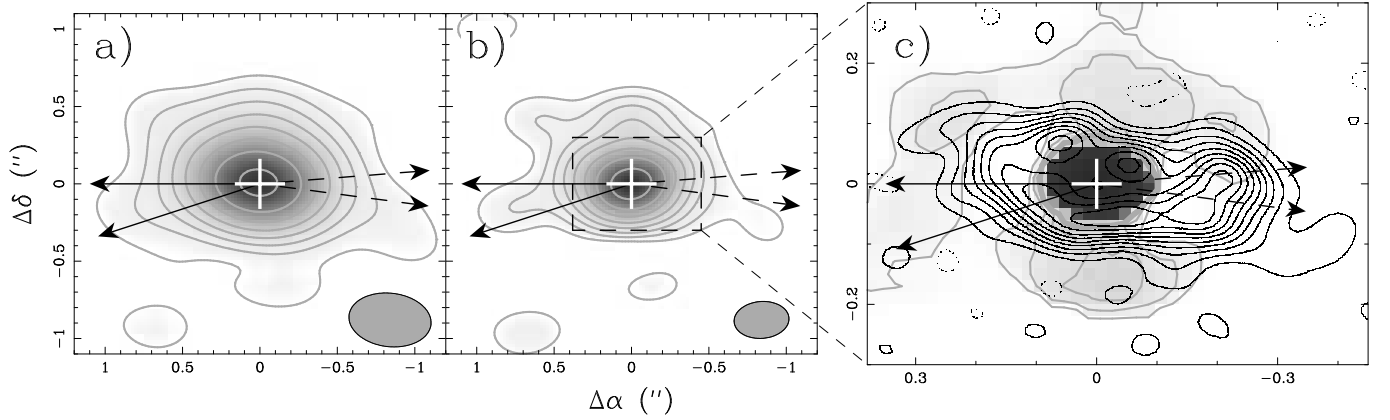

\centering
\putfig{0.73}{270}{f1.ps}
\figcaption[]
{350 GHz continuum maps toward the source.  The cross marks the central star
position.  The solid and dashed arrows indicate the outflow axes on the
blueshifted and redshifted sides, respectively.  (a) and (b) show the maps
at angular resolutions of \arcsa{0}{53}$\times$\arcsa{0}{35} and
\arcsa{0}{36}$\times$\arcsa{0}{24}, respectively.  The contour levels are 8$\sigma
(1-r^n)/(1-r)$, where $r=1.55$, $n=1,2,3,..$, with
$\sigma=3.7$ \mJyb{} in (a) and
$\sigma=2.4 $ \mJyb{} in (b).  (c) shows the CLEAN component map of the
continuum on top of the 23 GHz map of the H II shell adopted from
\citet{Martin1993}. The contour levels start at 26 K with a step of 52 K.
The compact emission peak at the center has a brightness temperature of $\sim$ 800 K.
\label{fig:cont}
}
\end{figure}

\begin{figure} [!hbp]
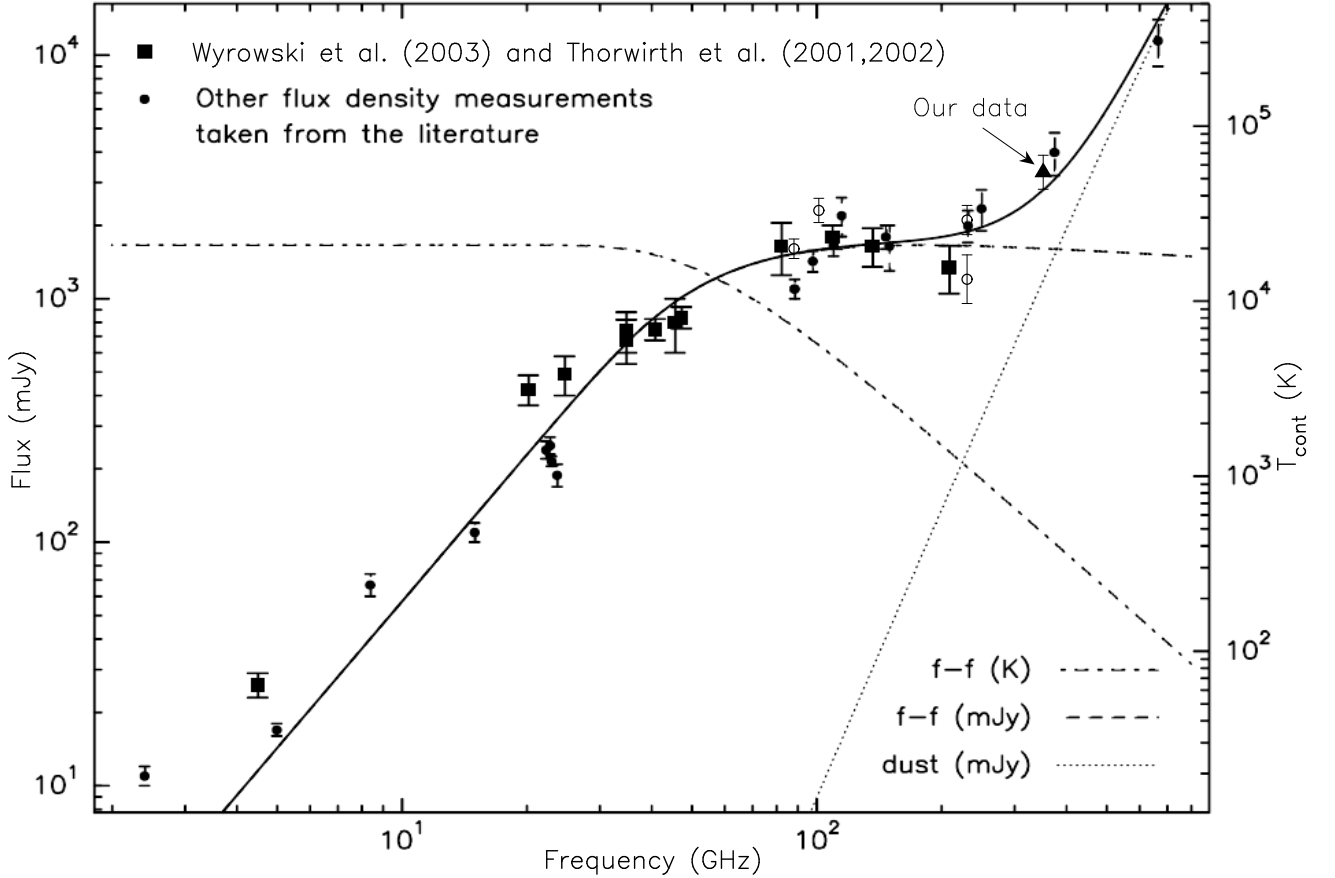

\centering
\putfig{0.73}{270}{f2.ps}
\figcaption[]
{Spectral energy distribution of CRL 618.  It is Figure 3 adopted from
\citet{Wyrowski2003}, and the fit in the figure was done by them.  For other
flux density measurements shown in the figure, please refer to that paper
for the references.  In their model, they assumed two components for the
continuum, a free-free (f-f) emission from the H II region and a thermal dust
emission from the envelope.  The brightness temperature of the H II region was derived assuming
a constant source size of $\sim$ \arcsa{0}{22}.  Previous OVRO (Owens Valley Radio
Observatory) data from \citet{SanchezS2004} and \citet{Sanchez2004} are
shown as open circles.  Our SMA data at 350 GHz is also included and its
flux density is consistent with the two-component model.  \label{fig:SED}
} \end{figure}

At 350 GHz, continuum emission is detected within \arcs{1} from the central
star elongated in the east-west direction along the main outflow axis (Fig. 
\ref{fig:cont}a), with a total flux density of $\sim$ $3.3\pm$0.7 Jy. 
According to previous model for the spectral energy distribution (SED) of
the source, the continuum emission at this frequency consists of two
components: free-free emission from the H II region near the central star
and thermal dust emission from the circumstellar envelope \cite[see Fig. 
\ref{fig:SED} and also][]{Wyrowski2003}.  Note that for the flux density in
the frequency between 80 and 360 GHz, \citet{Pardo2004,Pardo2007b} have
found that the total flux density from the lines represents less than
3\%-5\% of the continuum and thus will not affect the analysis of the SED. 
Our flux density is consistent with the previous model, arising from the two
components.  Since the emission detected here is within \arcs{1} from the
central star, the dust emission component here must be from the dense core
of the circumstellar envelope, which has an outer radius of $\sim$
\arcsa{1}{2} \citep{Sanchez2004}.

In order to distinguish the two components, we zoom into the emission peak
at the center at higher resolution in Fig.  \ref{fig:cont}b.  However, the
emission peak there is still not resolved.  Since it is detected with a S/N
ratio of more than 100, the structure there can be studied with the CLEAN
component map shown in Figure \ref{fig:cont}c.  In the map, a bright compact
emission peak is seen at the center inside the H II shell detected at 23 GHz
in the year of 1990 \citep{Martin1993}.  It has a brightness temperature of
$\sim$ 800 K, but the actual value must be higher because it is unresolved. 
It has a flux density of $\sim$ 1.4$\pm0.3$ Jy, roughly the same as that of
free-free emission required to fit the SED of the continuum source (see Fig. 
\ref{fig:SED}).  As a result, both the brightness temperature and flux
density indicate that it traces the H II region.  Note that the 23 GHz
continuum map has been shifted by $\sim$ \arcsa{0}{2} to the north in order
to match the center of the H II shell to the compact emission peak in our
map.  This position shift, if real, could be due to a proper motion of
$\sim$ 40 \vkm{} to the north.  As argued by \citet{Martin1993}, the H II
region is a filled region.  Since the H II region has a turnover frequency
$<$ 100 GHz (see Fig.  \ref{fig:SED}), the free-free emission at 350 GHz is
optically thin.  It appears as a peak at the center, indicating a presence
of a dense inner part there.  At 23 GHz, the H II region is optically thick. 
It appears as a shell probably because of an increase of electron
temperature \citep{Martin1993} or an increase of density \citep{Kwok1984}
there.  At 350 GHz, the shell is optically thin and it is not detected here
due to its not enough column density.

In the CLEAN component map, two faint emission peaks, one in the north and
one in the south, are seen at a radius of $\sim$ \arcsa{0}{14} ($\sim$ 126
AU) from the central star roughly in the equatorial plane perpendicular to
the outflow axes, surrounding the H II shell.  The peak in the north also
extends to the east and west surrounding the H II shell.  These
morphological relationships clearly indicate that these emissions trace the
limb-brightened edges of the innermost part of the dense core around the H
II shell.  The two emission peaks may arise from a density enhancement in
the dense core in the equatorial plane that helps confining the H II region
into a bipolar morphology.  The radius of the two emission peaks can
set an upper limit for the current radius of the H II region in the
equatorial plane.  Less emission is seen in the outflow axes, suggesting
that the dense core material there is cleared by the outflow.

%Thus, the H II shell may have expanded over the past 20 years for a
%few \arcsa{0}{01}.

Now it is clear that the extended emission in Figs. \ref{fig:cont}a and
\ref{fig:cont}b traces the dense core.  Since the free-free emission of the
H II region has a flux density of $\sim$ 1.4$\pm0.3$ Jy as discussed above,
the thermal dust emission from the dense core has a flux density of $\sim$
1.9$\pm0.4$ Jy.  In Figs.  \ref{fig:cont}b, the emission in the east is
resolved, extending to the northeast and southeast from the central star
(Fig.  \ref{fig:cont}b), likely tracing the dense core material around the
outflow cavity walls.  The emission is also seen extending $\sim$
\arcsa{0}{6} to the north from the central star, tracing the dense core that
may have a density enhancement perpendicular to the outflow axes.  However,
no counterpart is seen extending to the south.

\subsection{Molecular lines: Dense Core}

\subsubsection{Spectra}

Figure \ref{fig:spec} shows the spectra toward the inner region averaged
over a circular region with a diameter of \arcsa{0}{5}, from 342 to 346 GHz
and from 354 to 358 GHz.  The observed frequency has been converted to the
rest frequency using the systemic velocity of $-21.5$ \vkm{} as found in
\citet{Sanchez2004}.  Many molecular lines are detected, as listed in Table
1.  Most of them are from \nHC3N{} and its isotopologues, arising from
rotational transitions at various vibrational states, as found at lower
frequencies \citep{Wyrowski2003,Pardo2007b}.  As can be seen below, these
molecules trace mainly the dense core.  Most of their lines are isolated or
almost isolated, so that their line peak brightness temperature $\lpeak$ and
FWHM linewidth $\triangle v$ can be measured, as listed in Table 2, allowing
us to derive the properties of the dense core.  With an upper energy level
ranging from $\sim$ 300 K up to 2000 K, these lines can be used to probe the
properties of the dense core from the outer part down to the very inner part
enclosing the H II region.  Note that, sharp absorption dips are seen at
$\sim$ $-$16 \vkm{} in strong molecular lines, e.g., CO (deepest at $-$15.8
\vkm), CS ($-$15.8 \vkm), HCO$^+$ ($-$15.5 \vkm), HCN $v=0$ ($-$15.8 \vkm),
and H$^{13}$CN $v=0$ ($-$15.5 \vkm), due to an absorption by the extended
halo, which is cold and expanding at that velocity \citep{Sanchez2004}.  In
this paper, we study the dense core mainly with \nHC3N{} and its
isotopologues.  Other molecules trace mainly the outflow and will be studied
in another paper.

\subsubsection{Morphology}

%and singly substitute of $^{14}$N with $^{15}$N

The integrated intensity maps of the isolated and almost isolated lines of
\nHC3N{} and its isotopologues are shown in Figures \ref{fig:HC3N} \&
\ref{fig:HC3Niso}, respectively, in the order of increasing upper energy
level of the lines.  The figures show that for a given molecule, the
structure of the emission shrinks closer to the central star as we go to the
line with higher upper energy level, indicating that the temperature of the
dense core increases toward the central star.  In addition, comparing the
two figures, we can also see that, for a given similar upper energy level,
the lines of the isotopologues trace closer and thus warmer material than
the \nHC3N{} lines.  This is because the isotopologues are less abundant and
thus their lines are optically thinner.  For the isotopologues with doubly
substitutes of $^{12}$C with $^{13}$C, their abundances are very low, and
their lines with low upper energy level mainly trace the dense core close to
the central star (Fig.  \ref{fig:HC3Niso}).

For \nHC3N and its singly $^{13}$C substituted isotopologues, the lines with
the lowest upper energy level trace the outer part of the dense core that
can be resolved in our observations.  Figure \ref{fig:HC3N_cont} shows the
maps for two of these lines, one in \nHC3N{} and one in its isotopologue,
\H13CCCN{}, at a slightly higher angular resolution on top of the continuum
map.  Although the two lines have a similar upper energy level, the line of
the isotopologue traces closer to the central star as discussed above.  In
these maps, two emission peaks, one in the north and one in the south of the
central star, are seen surrounding the continuum emission peak, tracing the two
limb-brightened edges of the dense core in the outer part.  This
two-emission peak structure was also seen in a lower excitation line of
\nHC3N{} before, and was used to suggest an equatorial density enhancement
(mimicking a torus-like structure) in the dense core further out
\citep{Sanchez2004}.

The dense core is evacuated by the outflow lobes, with the emission around
the outflow cavity walls extending to the northeast, southeast, northwest,
and southwest, but with less emission along the outflow axes.  In addition,
the SE and SW outflow lobes evacuate more the southern part of the dense
core, reducing more the emission there near the central star, as seen in
\H13CCCN{}.  The major axis of the dense core, defined as the axis passing
through the two emission peaks and the central star position, has a position
angle of $\sim$ 3\degree{}, similar to that found by \citet{Sanchez2004},
almost perpendicular to the east-west pair (i.e., the major pair) of the
outflow lobes.  Thus, the dense core is likely to be perpendicular to that
pair of outflow lobes and is thus assumed to have an inclination angle of
$\sim$ 30\degree{} \citep{Sanchez2004}, with the nearside tilted to the west
and farside to the east.  Note that the outer part of the dense core is
expected to show a tilted ringlike structure in the maps, here we see more
like a ``C" structure because the emission is fainter in the western side of
the central star, due to a self-absorption to be discussed later.

Quite a few lines are also detected in HCN and its isotopologues (Fig. 
\ref{fig:spec}).  Some of them also trace mainly the dense core and the maps
of the isolated ones are shown in Figure \ref{fig:HCN_Vinyl}.  The HCN lines
at the vibrational states $v_2=1$ and $2$ trace the dense core because of
their high upper energy level and thus low number density at low
temperature.  The HCN line at ground vibrational state traces the outflow
and is thus not shown here.  The lines of H$^{13}$CN $v=0$ and $v_2=1$, and
HC$^{15}$N $v=0$ trace the dense core due to their low abundances.  Like
that of \nHC3N{}, the line with higher upper energy level traces the inner
part of the dense core.  We also detect many \C2H3CN{} (Vinyl Cyanide)
lines.  A total of 120 lines of this molecule have been detected at
lower frequency from 80 to 270 GHz by \citet[][see their Table
2]{Pardo2007b}.  In our frequency ranges, there are 12 lines with the line
strength $S_{ij} > 96$ (a factor of 3 is included here for the
spin-statistical weight of N).  Here, 11 of them are detected and 1 at
$\sim$ 356.247 GHz is lost in a strong \nHC3N line (Fig.  \ref{fig:spec}). 
These lines are weak, and thus we combine all of them to produce a map with
a high signal to noise ratio, as shown in Figure \ref{fig:HCN_Vinyl}.  It is
clear from the figure that these lines trace the innermost part of the dense
core due to the low abundance of the molecule.  This molecule could result
from the interaction of C$_2$H$_4$ and CN, as discussed in
\citet{Cernicharo2004}.

\subsubsection{Kinematics}

The kinematics of the dense core can be studied with the position-velocity
(PV) diagrams using the same two lines that show the resolved structure of
the dense core, as before.  The axial PV diagrams, with the cut
perpendicular to the dense core, show that the east side (or farside) is
mainly redshifted and the west side (or nearside) is blueshifted (see Figs. 
\ref{fig:pvs_env}a, b), in opposite to that seen for the outflow.  Note that
for \nHC3N{}, the blueshifted emission with the velocity $\lesssim$ $-$10
\vkm{} should be ignored because it is contaminated by another weak \nHC3N{}
line centered at $-$20 \vkm{}.  This PV structure clearly supports that the
core is expanding away from the central star.  Negative contours are seen on
the blueshifted side, due to an absorption of the continuum emission and the
line emission by a cold layer on the nearside.  Thus, the dense core appears
fainter in the west of the central star, as seen above in Figure
\ref{fig:HC3N_cont}.  The expansion velocity in each emission line is
proportional to the maximum velocity, either redshifted or blueshifted. 
Since the blueshifted side is self-absorbed, the redshifted side is used,
and the redshifted velocity is higher in the \nHC3N{} line than in the
\H13CCCN{} line.  Since the \nHC3N{} line traces outer region than the
\H13CCCN{} line, this suggests that the expansion velocity increases with
the distance from the central star.  The equatorial PV diagrams, with the
cut along the major axis of the dense core, show an incomplete ringlike PV
structure due to the absorption on the blueshifted side (Figs. 
\ref{fig:pvs_env}c, d).  This ringlike PV structure indicates that for a
given position offset from the central star, the blueshifted and redshifted
emission are seen.  This is expected because the dense core has a small
inclination angle and it is thick enough for the cut to pass through both
the farside and nearside of the dense core.

As mentioned above, the expansion velocity is found to increase with the
distance from the central star.  Here we can study quantitatively how fast the
increase is, using the linewidth and the angular radius of the dense core
seen in the lines of \nHC3N and its isotopologues.  We first measure the
angular diameter and then divide it by two to obtain the angular radius. 
The angular diameter of the dense core in different line emission can be
defined as the full extent in the major axis at the half maximum of the
emission peak.  It can be measured from those integrated maps (Figs. 
\ref{fig:HC3N} \& \ref{fig:HC3Niso}) that have enough signal to noise ratio,
as listed in Table \ref{tab:measure}.  Figure \ref{fig:emis_size} shows the FWHM linewidth,
$\triangle v$, versus the angular radius $r$ of the dense core.  It shows
that the linewidth and thus the expansion velocity
increases roughly linearly with the angular radius.  In
this figure, we exclude the lines with the highest upper energy level, due
to their low signal to noise ratio.  Also, we exclude the zero vibrational
line of \nHC3N{}, which could be affected by the outflow lobes.

\subsubsection{Physical Properties}

Population diagram can be used to estimate the mean excitation temperature
and the column density of \nHC3N{} toward the inner part of the dense core. 
It is a diagram that plots the column density per statistical weight in the
upper energy state in the optically thin limit, $N_u^\textrm{\scriptsize
thin}/g_u$, versus the upper energy level $E_u$ of the \nHC3N{} lines (Fig. 
\ref{fig:rottemp}).  Here $N_u^\textrm{\scriptsize thin}=(8\pi k\nu^2/hc^3
A_{ul}) W$, where the integrated line intensity $W=\int T_B dv \approx 1.06
T_B^p \cdot \triangle v$.  The lines of the singly $^{13}$C substituted
isotopologues can also be included in the diagram once their integrated line
intensity is multiplied by the abundance ratio of \nHC3N{} to the
isotopologues.  The abundance ratio has been found to be $\sim$ 10
\citep{Wyrowski2003}.  This ratio is consistent with our observations
because the line intensity of the isotopologues, after multiplied by this
ratio, becomes aligned with that of the \nHC3N{} lines in the diagram.  The
diagram shows almost a straight line if we exclude the data points for the
\nHC3N{} lines with $E_u/k < 1000$ K.  Those data points deviate
significantly from the straight line because for those data points, (1) the
lines become optically thick and (2) a significant fraction of the emission
is outside the region that is used to derive the line intensity.  The data
points lie almost in a straight line, suggesting that the lines are mostly
optically thin and arise from LTE material.  Fitting the data with a
straight line, we find that the mean excitation temperature and column
density of the \nHC3N{} molecules are $\sim$ 350 K and 9$\times 10^{16}$
\cms{}, respectively.  The column density should be considered as a lower
limit, because (1) the line of the isotopologues at the lowest upper energy
level is not optically thin, showing an absorption dip in the blueshifted
velocity as discussed above, and (2) the emission for line at lower upper
energy is further away from the central star.

\section{Model}

In order to derive the properties of the dense core more accurately, we
construct a radiative transfer model to calculate the free-free emission of
the H II region, the thermal dust emission and molecular line emission of
the dense core to compare with the observations.

Figure \ref{fig:modelsketch} shows a schematic diagram for our model.
As discussed in Sec.  \ref{sec:cont}, the H II region is a filled region at the center
elongated in the east-west direction.  The radius of the H II region in the
equatorial plane is uncertain and it might have grown to $\sim$
\arcsa{0}{14} as discussed.  Thus, the H II region is assumed to be an
ellipsoid with a size of
\arcsa{0}{6}$\times$\arcsa{0}{28}$\times$\arcsa{0}{28} elongated in the
east-west direction, which is an approximate representation of the size and
morphology of the 23 GHz continuum map (Fig.  \ref{fig:cont}).  The inner
radius of this H II region is unresolved and set to a small value of
\arcsa{0}{01} (9 AU), which is more than 20 times smaller than our
resolution and thus should not affect our model comparison.  In order to
produce the bright compact emission peak at the center, the electron density
in the H II region is assumed to decrease with the radial distance from the
central star, $r$, as follows \begin{equation} n_e=n_{e0}
(\frac{r}{\arcsaq{0}{05}})^{-2} \;\textrm{cm}^{-3} \end{equation} The
electron temperature of the H II region is assumed to be constant at 15,000
K, in between that derived by \citet{Martin1993} and \citet{Wyrowski2003}. 
Note that the H II shell detected at 23 GHz could suggest an increase of
electron temperature \citep{Martin1993} or density \citep{Kwok1984} in the
outer part of the H II region.  However, at 350 GHz, the shell is optically
thin and it is not detected here because of its low column density. 
Therefore, its flux can be ignored as compared to that of the dense H II
core at the center.  As a result, the possible increase in the electron
temperature or/and density is not included in our model.

The dense core is assumed to be spherical originally, with the inner radius
set by the outer boundary of the H II region and the outer radius set to
\arcsa{1}{2}, as found in \citet{Sanchez2004}.  The dense core is excavated
by the outflow lobes.  For simplicity, we assume two outflow cavities, one
in the east and one in the west, both with an half opening angle of
25\degree{} (Figure \ref{fig:modelsketch}), as judged from the \nHC3N map in
Fig \ref{fig:HC3N_cont}a, which shows emission extending to the northeast,
northwest, southeast, and southwest.

The dense core is dusty and molecular, producing both the thermal dust
emission and molecular emission.  For simplicity, the dust and molecular gas
are assumed to have the same temperature.  This temperature was first
assumed to have a single power-law distribution as follows
\begin{equation} 
T \propto r^{-p} \;\;\textrm{K} 
\end{equation}
We found that when $p=1.8$, this
temperature distribution can roughly reproduce the low excitation lines in
the outer part.  However, it can not produce enough emission for the high
excitation lines in the inner core, because the temperature there was
too high.  Therefore, the temperature is assumed to have the following
two power-law distribution with a turning point at $\ro$

\begin{eqnarray}
T(r)= \left\{ 
\begin{array}{cl}
T_0\big(\frac{r}{\ro}\big)^{-p_i} & \;\;\textrm{if}\;\; r < \ro, \\
T_0\big(\frac{r}{\ro}\big)^{-p_o} & \;\;\textrm{if}\;\; r \geq \ro
\end{array}
\right. 
\end{eqnarray} 
\noindent with the power-law index $p_i < p_o$.
In our model, the expansion velocity is assumed to increase
linearly with the radius, \begin{equation} \vexp= v_0 (\frac{r}{r_0})
\;\;\textrm{\vkm} \end{equation} as we discussed earlier.  Note that,
however, the dense core will have a maximum (i.e., terminal) expansion velocity, which is
assumed to be equal to the expansion velocity of the extended halo or $\sim$
16 \vkm{}, as found earlier.  The mass-loss rate in the dense core is
assumed to be constant for simplicity, as in \citet{Fonfria2011}.  Thus, the
number density of molecular hydrogen in the dense core becomes
\begin{equation} n = n_0 (\frac{r}{r_0})^{-3} \;\;\textrm{\cmc}
\end{equation}

%gas to dust ratio about 100
%dust opacity per dust mass about 1 cm^2/g
%==> 1/100 = 10^-2

For the thermal dust emission in the dense core, the dust opacity per unit
gas mass, $\kappa_\nu$, is assumed to be a free parameter in the order of
10$^{-2}$ cm$^2$ g$^{-1}$ at 350 GHz \citep{Sahai2011}.  In the model, the
molecules that trace the dense core are included.  The outflow lines, e.g.,
CO, CS, HCO$^+$, and HCN $v=0$, are excluded.  The abundance of \nHC3N{} is
assumed to be $2\times10^{-7}$ as found in \citet{Sanchez2004}.  The
abundances of other molecular species are derived from our model
by fitting the line profiles of each of these species.

Radiative transfer is used to calculate all the emissions with an assumption
of LTE.  The thermal linewidth and the linewidth due to a turbulence
velocity of 2 \vkm{} are included.  The systemic velocity is assumed to be
-21.5 \vkm{}.  Also, we rotate our model counterclockwise by a P.A.  of
3\degree{} and tilt it with an inclination of 30\degree{} to match the
observations.  The distance of the source is assumed to be 900 pc.

\subsection{Model Results} \label{sec:modelres}

%dense core in the
%350 GHz continuum and three emission lines.  

Figure \ref{fig:fitspec} shows the fit of the spectra with our simple model. 
The parameters are listed in Table \ref{tab:model}, and the profiles of the
temperature, density, and velocity in the dense core are shown in Figure
\ref{fig:modprofile}.  In order to produce the observed flux density for the
H II region, we obtain $n_{e0}=6.4\times10^{6}$ \cmc{}.  For the inner,
denser part of the H II region, the emission measure (EM) averaged over a
radius of \arcsa{0}{1} is $\sim$ 4.2$\times10^{10}$ cm$^{-6}$ pc, in
agreement with that derived by \citet{Wyrowski2003}, who assumed a radius of
$\sim$ \arcsa{0}{1} for the H II region.  

%For H$^{13}$CN,
%the abundance is derived from the $v=0$ and $v_2=1$ lines.  The optical
%depths averaged over the central region with a radius of \arcsa{0}{5} (where
%most of the emission arises) are $\sim$ 4.0 and 0.06 for $v=0$ and $v_2=1$
%lines, respectively.  Thus, the H$^{13}$CN abundance is strongly constrained
%by the $v_2=1$ line.

%  For HCN, the abundance is derived from the $v_2=1$ and
%$v_2=2$ lines.  The optical depths averaged over the central region with a
%radius of \arcsa{0}{35} (where most of the emission arises) are $\sim$ 0.70
%and 0.08 for $v_2=1$ and $v_2=2$ lines, respectively.  Thus, the HCN
%abundance is strongly constrained by the $v_2=2$ lines, and to a lesser
%extent by $v_2=1$ lines.  

With the assumed abundance of \nHC3N{}, we obtain the temperature,
density, and velocity distributions of the dense core by fitting all the
\nHC3N{} lines that trace the dense core, including both the optically thin
and optically thick lines.  Then we derive the abundances of other molecular
species by fitting their lines.  The \nHC3N{} isotopologues are much less
abundant than \nHC3N{} and their lines are mostly optically thin.  For
\C2H3CN{} and HC$^{15}$N, their lines are optically thin.  For H$^{13}$CN,
the abundance is strongly constrained by the $v2 = 1$ J=4-3 line (at
345238.7103 MHz see Table \ref{tab:linedetection}), as this is optically
thin ($\tau \sim 0.06$) in the region where the bulk of its emission arises
(radii $<$ \arcsa{0}{5}).  For HCN, the abundance is strongly constrained by
the v2 = 2 J=4-3 (at 356301.1780 MHz) and v2=1 J=4-3 (at 354460.4346 and
356255.5682 MHz) lines (see Table \ref{tab:linedetection}), as these have
optical depths less than unity ($\tau \sim$ 0.08 and 0.70, respectively) in
the region where the bulk of their emission arises (radii $<$
\arcsa{0}{35}).  Nonetheless, since our model assumes LTE, there could be
uncertainty in our calculation of the abundances because of non-LTE
effects.

%is probably not
%related to optical depth effects (as discussed) but maybe to the
%uncertaities of the v>0 level population relative to the v=0... 
%(excitation, non-LTE effects) }

%  Nonetheless, since \citet{Thorwirth2003} has
%detected direct l-doubling transition lines of HCN (for which the line
%strengths are very low), further work is needed to check if we could
%under-estimate the HCN abundance.}

It is clear from Figure \ref{fig:fitspec} that our model can roughly
reproduce the line peak and linewidth for most of the lines that trace the
dense core.  As mentioned early, the outflow lines are excluded in the
fitting.  Since they are strong and broad, the lines that are close to them
are significantly affected and thus can not be fitted well.  In order to
further check the reliability of our model, we also present the comparison
of the synthetic and observed 350 GHz continuum maps as well as the maps for
three emission lines.  Since \H13CCCN{} is mostly optically thin and bright,
we choose three of its isolated lines at the different vibrational states
that trace the different parts of the dense core.  As can be seen from
Figures \ref{fig:fitstr} and \ref{fig:fitPV}, our model can also roughly
reproduce the structures and the kinematics of the dense core in these
emissions.  Our model is symmetric and thus will not account for the
asymmetric structure.
%Future model will address the asymmetric structure when the inner part of
%the dense core is better resolved.

%if we assume the same abundance for them. 

Nonetheless, there is still some minor discrepancy between our model and the
observations.  For the two bright HCN $v_2=1$ lines, although our model can
roughly fit their line peak, it produces a linewidth smaller than the
observed.  This is probably because these two lines might have been affected
by the outflows, as found in the line in the ground vibrational state of
HCN.  Also, for the three \nHC3N{} isotopologues with doubly substitutes of
$^{13}$C, we expect to see one line for each of them in our frequency
ranges.  However, the line of H$^{13}$C$^{13}$CCN at 342467.9204 MHz is not
observed.  Moreover, although we can roughly reproduce the peak of the lines
for other two isotopologues, H$^{13}$CC$^{13}$CN and HC$^{13}$C$^{13}$CN,
but the lines in our model are too broad.  Since these lines are very weak,
future observation at higher sensitivity will be needed to confirm our result.

%Fonfria used TEXES spectrograph on the IRTF 3m telescope with a
%spectrograph slit \arcsa{1}{4} wide.  From Fonfria et al.  2011, ours
%observed dense core was divided to his central HII region (ac=0.135"
%spherical), Z1 (1.5ac=0.2"), Z2 (2ac=0.27"), and inner part of the Z3 (11ac
%~ 1.5") zones.  With C2H2, the (vibrational + rotational) temperature is
%550 K at the innermost of Z1 and decreases to 400 between Rc and R1, and
%reaching 250 K at R3.  With C4H2, the temperature is found to be 150 K
%higher.

%till $r=$ \arcsa{1}{2} 
%as derived from CO J=2-1 in \citet{Sanchez2004} at $\sim$   
%\arcs{1} angular resolution,
%similar to that found from fitting the upper energy level of the \nHC3N{}
%lines.  It is also 

%mean T = \int 4*pi*r^2*n*Tdr/\int 4*pi*r^2*n dr
%mean column density =\int 4*pi*X(HC3N)*n*dr/(pi*rout**2)

In our model, since the density and temperature both decrease rapidly with
the increasing radius, the line emissions are mainly arisen from the inner
part of the dense core within $\sim$ \arcsa{0}{7} from the central star, as
seen in the observations.  In this part of the dense core, the expansion
velocity increases from $\sim$ 3 \vkm{} to the maximum velocity of $\sim$ 16
\vkm{} at $r\sim$ \arcsa{0}{7} (630 AU).  The outer part of the dense core
does not change much the spectra, it only produces a deep absorption dip at
$\sim$ $-16$ \vkm{} for the $v=0$ lines of HC$^{15}$N, H$^{13}$CN, and even
\nHC3N.  This is because the expansion velocity there reaches and stays at
the maximum velocity of $\sim$ 16 \vkm{}.  

The parameters in our model are consistent with what we estimated earlier. 
For instance, the mean column density of \nHC3N{} averaged over a radius of
\arcsa{0}{5} is $\sim 2\times 10^{17}$ \cms, only about 2 times the lower
limit derived from the population diagram.  The temperature averaged over a
radius of \arcsa{0}{5} is $\sim$ 340 K, similar to the mean excitation
temperature derived from the population diagram.  The temperature power-law
index in the outer part of the dense core is similar to that found in
\citet{Sanchez2004} by fitting a low excitation line of \nHC3N{}.  The
temperature power-law index in the inner part of the dense core is the same
as that found in \citet{Wyrowski2003}.  The temperature within $r_0$ (i.e.,
\arcsa{0}{22}) here is also consistent with that found by
\citet{Fonfria2011}.  By fitting the infrared spectra of C$_2$H$_2$ and
C$_4$H$_2$, they found that the temperature decreases from $\sim$ 600 K to
400 K from the innermost part of the dense core to $r_0$, similar to our
model.

Previously, a sophisticated model has been proposed by Pardo et al. 
\citep{Pardo2004,Pardo2005,Pardo2007a,Pardo2007b} to explain the various
molecular emissions of CRL 618 observed with the IRAM 30m telescope in the
frequency range from 80 to 276 GHz.  The dense core here can be considered
as a refined version of the slowly expanding envelope (SEE) in their model. 
In their model, the SEE has an outer radius of \arcsa{1}{5}, slightly larger
than that of the dense core.  It has a temperature of 250-275 K, slightly
lower than the mean temperature in the dense core.  The expansion velocity
field has a radial component ranging from 5 to 12 \vkm{}, with a possible
extra azimuthal component reaching 6 \vkm{} at \arcsa{1}{5}, and is thus not
much different from that in the dense core.  The column density of HC$_3$N
is $(2.0-3.5)\times10^{17}$ \cms{}, also similar to that found in the dense
core.  In their model, the SEE is surrounded by a colder ($\sim$ 60 K) and
outer (a radius from \arcsa{1}{5} to \arcsa{2}{25}) circumstellar shell
(CCS) created during the AGB phase, responsible for most of the rotational
emission from HC$_3$N $\nu_7$ and $v=0$ and HC$_5$N $v=0$ \citep{Pardo2005}. 
In our model, there is no need for such an extended shell because our
observations mostly probe the central region within $\sim$ \arcsa{0}{7}
from the central star.

%mass-loss rate
%without He
%4*pi*0.2**2*(2*1.66e-24*5.3e8)*4.5e5*3.156e7*(900*1.5e13)**2/1.989e33
%if He included, then 1.1 times

%total mass
%without He
%4*pi*0.2**3*(2*1.66e-24*5.3e8)*log(1.2/0.14)*(900*1.5e13)**3/1.989e33
%if He included, then 1.1 times the value

% For each opening with an half opening angle of 30degree
% \int dOmege = 2\pi \int sin \theta d\theta 
%             = 2\pi [-cos \theta]_0^30degre
% So for a pair of opening, we have \dOmega = 4 \pi (0.1339)
% This opening will move about 13.4% of the mass if we assume the core
% has an inner radius of 0.1"
% However, the inner radius of the core is not uniform because of the 
% the H II region, which has a hole 0.2"x0.6" in diameter.
% Thus, this H II region will remove some of the mass too.
% However, since most part of it is already removed by the opening,
% We can ignore this part.
% In math
% h= \int_0^a R dR \int_zin^zout ((R^2+z^2)/ro^2)^(-3/2) dz
% ro=0.2
% f=((R^2+z^2)/ro^2)^(-3/2)
% g=R*Integrate[f,z]
% z=0.1*Sqrt[1-R^2/0.1^2] ==> zin
% gi=g
% z=0.3*Sqrt[1-R^2/0.1^2] ==> zout
% go=g
% g=(go-gi)
% h=Integrate[g,{R,0,0.1}] = 0.0018314

\section{Discussion}

Our model is very simple. Nonetheless, it already can produce a reasonable
fit to the observations of the dense core.  The abundance for each species
is assumed to be constant, and no chemical effect is included.  The dense
core is assumed to be in LTE, which may not be the case near the central
star because of a possible infrared pumping there.  The different power-law
index for the temperature in the inner part of the dense core could be
related to this.  The dense core is assumed to be spherical with 2 conical
cavities.  The actual dense core could have a density enhancement in the
equatorial plane, as hinted in the continuum map and the map of the low
excitation line of \nHC3N \citep{Sanchez2004}.  This density enhancement in
the equatorial plane, if real, could help confining the H II region into a
bipolar morphology.  Future model including these effects will be needed for
a more detailed comparison.

%which is a factor of 5
%higher than that found in \citet{Sanchez2004}.

The dense core could result from a recent heavy mass-loss episode that ends
the AGB phase.  The mass-loss rate in the dense core is $\dot{M} = 4\pi r_0^2
n_0 v_0 \mH2 \sim 1.15\times10^{-3}$ \smpy.  It is $\gtrsim$ 2 orders of magnitude
higher than the typical values of the AGB wind, indicating that the dense
core could arise from an enhanced heavy mass loss that ends the AGB phase
\citep{Huggins2007}.  The total mass in the dense core is $\sim$ 0.47
$\solarM$.  
If we divide this mass by the mass-loss rate, we will have a
dynamical age of $\sim$ 400 yr for the dense core.
%The time elapse from the innermost radius of \arcsa{0}{1} to \arcsa{1}{2} is 
%$\int \frac{ds}{\vexp} \sim 460$ yrs.

%mass-loss rate of the stellar wind
%4*pi*(0.05*900*1.5e13)**2*6.4e6*1.66e-24*3e7*3.156e7/2e33

%mass of the compact HII peak within \arcsa{0}{1} is
%4*pi*(0.05*900*1.5e13)**2*6.4e6*1.66e-24*(0.1*900*1.5e13)/2e33

\subsection{Nature of the H II region}

%wind could be isotropic with the density decreasing with $r^{-2}$ in all
%directions, as assumed in our model.

%In our model, the region that gives rise to this emission peak has an outer
%radius of $\sim$ \arcsa{0}{1} and thus a mass of $\sim 4\times 10^{-5}$
%$\solarM$.
%The H II region trace an ionized stellar wind 

At low frequency at 23 GHz, the H II region appears as a shell structure. 
Two scenarios were proposed to explain this shell structure.  One scenario
suggested that the H II region traces an ionized stellar wind photoionized
by the central star \citep{Martin1988,Martin1993}.  In this scenario, the H
II region is a filled region and it appears as a shell due to a rapid
increase of the electron temperature toward the edge.  To have a filled
region, the central star is required to be still in the mass-loss phase
\citep{Martin1988}.  The other scenario suggested that the H II shell
represents a nascent PN shell or a contact discontinuity as produced in an
interacting-stellar winds model \citep{Kwok1984}.  In this model, a new fast
wind generated in the PN phase interacts with the envelope formed by the
stellar wind in the AGB phase.  In both scenarios, a wind is required to be
ejected from the central star.

Those two scenarios were proposed before the detections of fast moving
optical collimated outflow lobes \citep{Trammell2002,Sanchez2002} and fast
moving massive molecular outflows \citep{Sanchez2004} in CRL 618.  It is now
believed that these fast moving phenomena are produced by a fast
collimated post-AGB wind ejected after the AGB phase (or earlier at the end
of the AGB phase) \cite[see e.g.][]{Lee2003,Sanchez2004}.  This fast wind
interacts with the dense core, also producing a contact discontinuity at the
interface.  Thus, the H II shell may trace this contact discontinuity
photoionized by the central star.  

The compact H II peak at the center within the H II shell is seen for the
first time in CRL 618.  It may trace the post-AGB wind at the base
photoionized by the central star.  On the other hand, since the central star
has become hot ($\sim$ 30,000 K) and luminous ($\sim$ 10$^4$ $\solarL$), a
fast isotropic ionized wind may also be ejected from the central star by
radiation pressure.  Therefore, the compact H II peak may trace this fast
wind as well.  Further observations at higher resolutions are really needed
to resolve it in order to check these possibilities

%Fast wind mass-loss rate
%6.4e6*1.66e-24*(0.05*900*1.5e13)**2*1e8*3.156e7/2e33*4*pi = 9.6e-5 solar mass per yr.

%Contact discontinuity also expands

%new fast wind through radiation pressure acting directly on the gas ==> isotropic
% 1 or 2 orders of magnitute greater than the AGB velocities

%The maximum luminosity of a source in hydrostatic equilibrium is the
%Eddington luminosity.  If the luminosity exceeds the Eddington limit, then
%the radiation pressure drives an outflow.

%For pure ionized hydrogen
%LEdd=3.2e4 (M/Ms)Ls
% CRL 618 1.22e4 *Ls (D/kpc)^2 = 1e4 Ls for D=0.9kpc

%The dynamical age of
%the central H II peak is only $\sim$ 1.5 yrs.

\subsection{Dust Properties}

%As argued in \citet{Sahai2011}, this value of dust
%opacity at $\lambda=0.86$ mm (350 GHz) suggests 

%The dust opacity per gas mass is 0.022 cm$^2$ g$^{-1}$. The gas to dust
%ratio in the AGB wind and superwind is uncertain and could be in a range of
%50-200.  Thus, the dust opacity per dust mass is 1.1-4.4 cm$^2$ g$^{-1}$.

As discussed in Sec. \ref{sec:cont}, at 350 GHz, the extended continuum emission
traces the dust emission from the dense core and it has a flux density of
$\sim$ 1.9$\pm0.4$ Jy.  Previously in single-dish observations, continuum
emission was detected with a flux of $\sim$ 12$\pm2.5$ Jy at 670 GHz (450
\micron{}, see also Fig.  \ref{fig:SED}) and 23$\pm4$ Jy at 850 GHz (350
\micron{}) \citep{Knapp1993}.  As argued by the authors, the continuum
emission at those two frequencies is highly dominated by the dust emission
of the circumstellar envelope and thus the fluxes there can be considered as
the upper limits for the dust emission in the dense core.  Fitting to the
fluxes at the 3 frequencies, we find the flux of the dense core $\propto
\nu^{2.8\pm0.2}$, which results in a dust emissivity index
$\beta=0.8\pm0.2$, as an upper limit.  This value of $\beta$ is in good
agreement with that derived by \citet{Knapp1993} for the circumstellar
envelopes around five highly evolved stars, including CRL 618.  A value of
$\beta \lesssim 1$ has been used to imply a presence of large (mm-sized)
grains in protoplanetary disks \citep{Draine2006}, as well as in torii and
disks around post-AGB stars \citep{Sahai2011}.  Thus, there could be large
(mm-sized) grains in the dense core of CRL 618 as well down to $\sim$ 126 AU
(\arcsa{0}{14}) from the source.

In the dense core, the dust opacity per gas mass is found to be $\sim$
0.022$\pm0.004$ cm$^2$ g$^{-1}$ in our model.  The gas-to-dust ratio is
uncertain.  It was found to be $\sim$ 63 with single-dish observations,
averaged over both the extended halo and the dense core \citep{Knapp1993}. 
If we assume this ratio for the dense core, then the dust opacity per dust
mass will be $\sim$ 1.4$\pm0.3$ cm$^2$ g$^{-1}$, the same as that adopted to
derive the mass of the disks and torii around the post-AGB stars
\citep{Sahai2011}.  Note that the gas-to-dust ratio could be a factor of 2
larger in the dense core as compared to that in the extended halo
\citep{Meixner2004}, and so could be the dust opacity.

As discussed above, the dense core could result from a heavy mass loss at
the end of the AGB phase.  The mass loss (or wind) could be driven by
radiation pressure of the stellar light on dust grains.  For radiation
driven wind, most of the acceleration is believed to take place in a very
thin innermost part where the dust grains have sizes of up to a micrometer. 
Here in CRL 618, however, we see that the acceleration continues out to
\arcsa{0}{7} (630 AU) even though the dust grains could have grown to mm
sizes.  Thus, further observations are needed to study the cause of this
acceleration in the dense core.

%Outflow interaction?

\subsection{Isotope Ratios}

%Wyrowski et 22", 17.6 and 11.5" at 109, 136 and 209 GHz, us IRAM 30m
% Pardo also IRAM 30
% Pardo 2005, SSE 1.5", CCS from 1.5" to 2.25"

% As argued in \citet{Sanchez2004}, since most
%the \nHC3N molecules are in the dense core,

%The determination of isotopic abundances plays an important
%role in our understanding of nucleosynthesis in evolved stars.
%For the stars with masses in the range $2.5 <= M/Ms <= 6$,
%the hot bottom burning (HBB) may occur during the AGB stage
%and induce 12C/13C to further decrease to 3.5 (Frost et al.
%1998).

%Frost, C. A., Cannon, R. C., Lattanzio, J. C., Wood, P. R., & Forestini, M. 1998,
%A&A, 332, L17

% The
%decrease of the isotopic ratio toward the center may also be seen in the
%well-known C-rich AGB star IRC+10216.  In that source, the isotopic ratio
%was found to be $\sim$ 45 in the circumstellar envelope with single-dish
%observations \citep{Cernicharo2000}.  It may decrease to 10-15 in the inner
%part of the envelope, as found by \citet{Patel2011} with the SMA
%observations, although more observations are needed for confirmation.  
%using the Arizona Radio Observatory Sub-millimeter Telescope at $\lambda=1$
%mm by \citet{Milam2009},

%local ISM value ($\sim$ 68). 

\subsubsection{Carbon}

Isotopic ratios can be used to constrain current nucleosynthesis models in
evolved stars.  Previously with the IRAM 30m single-dish observations,
\citet{Pardo2007a} found that the isotopic ratio of $^{12}$C/$^{13}$C is 15
in the dense core [or slowly expanding envelope (SEE) in their model] and
$\gtrsim 40$ in the extended halo [or circumstellar shell (CSS) in their
model], using e.g., \nHC3N, HCN, HNC, and their isotopologues in lower
transition lines.  Here, with the observations at higher angular resolutions
in higher transition lines, the isotopic ratio of $^{12}$C/$^{13}$C is found
to be $\sim$ 10$\pm3$, using \nHC3N and its isotopologues.  This value is
the same as that found by \citet{Wyrowski2003} and similar to that found by
\citet{Pardo2007a} using the same species in lower transition lines.  This
is expected because this species is mainly present in the dense core
\citep{Sanchez2004}.  The ratio is found to be $\sim$ 8$\pm3$, using HCN and
its isotopologues.  Thus, the mean value of the ratio is $\sim$ 9$\pm4$,
similar to that found by \citet{Pardo2007a} in the dense core using various
molecules.  Note that lower $^{12}$C/$^{13}$C ratios have been seen before
in C-rich PPN/PN, for instance, a value of $\sim$ 5 in Boomerang Nebula
(PPN) \citep{Sahai1997}, and a value of $\sim$ 3 in M1-16 [which is a very
young multipolar PN with dense core (so very similar to CRL 618) and compact
H II region] \citep{Sahai1994}.

Comparing to a recent extensive study by \citet{Milam2009} with ARO
single-dish observations in CN, CO and their isotopologues, we find that our
value is smaller than those found in the circumstellar envelopes around
C-rich stars, which are $\sim$ 25-90, but it is in the lower limit of those
found in the circumstellar envelopes around O-rich stars, which are $\sim$
10-35.  Note that in their study, the ratios are the values averaged over a
large extent of the circumstellar envelopes including both the extended
halos and dense cores.  For CRL 618, they found a ratio of $\gtrsim$ 32,
much larger than our value, probably due to a high ratio of $\gtrsim$ 40
found in the extended halo \citep{Pardo2007a}.  In this case,
high-resolution observations are really needed to derive the ratio in the
dense core.

More recently, by modeling Herschel data of CO/$^{13}$CO and HCN/H$^{13}$CN
lines in CRL 618, \citet{Wesson2010} found a  $^{12}$C/$^{13}$C ratio of 21, which is
intermediate between the high value of $\gtrsim$ 40 and our low value of 9. 
The region they probed has a temperature of 70-230 K, and thus corresponds
to our dense core in the middle part from $\sim$ \arcsa{0}{3} to
\arcsa{0}{6}.  As a result, the $^{12}$C/$^{13}$C ratio indeed appears to
decrease (perhaps in a continuous manner) from $\gtrsim$ 40 in the extended
halo to 9 in the inner part of the dense core, as argued by
\citet{Pardo2007a}.  The time scale for this change is short.  In
\citet{Pardo2007a}, the ratio of $\gtrsim$ 40 was obtained at a radius of
$\sim$ \arcs{2}.  Our value is mostly from the inner part and can be assumed
to be at a radius of $\sim$ \arcsa{0}{3}.  Assuming an expansion velocity of
$\sim$ 16 \vkm{}, then the time scale is only $\sim$ 450 yrs.

%you should also reference Wesson et al. 2010 A&A 518, L144, who find 
%12C/13C of 21 modeling Herschel data of CO/13CO and HCN/H13CN lines. 
%this is intermediate between the high value of 40 and low value of 9 and 
%suggests that the 12C/13C ratio has been decreasing (perhaps in a 
%continuous manner) over the mass-loss time-scale encompassed by the 
%large-beam observations, i.e. ~8600 yr at a radius of ~30 arcsec [from 
%the Milam et al. CN N=1-0 data] with vexp=15.5 kms/s, and the SMA data 
%(170 yr). make an explicit statement about this -- namely that the 
%12C/13C ratio appears to have changed dramatically over this time-scale, 
%and that there is currently nucleosynthesis model can explain this result.

Our $^{12}$C/$^{13}$C ratio in the dense core is much smaller than the solar
value, which is $\sim$ 89.  For an AGB star, the $^{12}$C/$^{13}$C ratio is
first expected to go down from the solar value due to the first dredge-up. 
This is because $^{13}$C, which is produced in CNO cycle during the RGB
phase, is transported out to the envelope by the first dredge-up.  Then when
the 3rd dredge-up occurs adding fresh $^{12}$C to the envelope, the
$^{12}$C/$^{13}$C ratio starts going up again.  When the envelope becomes
C-rich, this ratio is expected to be $\gtrsim$ 35, which is what is seen in
the extended halo in CRL 618.  However, it is unclear how the ratio can
decrease again after the envelope has become C-rich.  As suggested by
\citet{Pardo2007a}, one possibility is to have a late CNO cycling phase that
follows He burning phase, as in Sakurai's object \citep{Asplund1999}.  Note
that the $^{12}$C/$^{13}$C equilibrium value from CNO cycling is $\sim$ 3.5
\citep{Asplund1999}.

%Another possibility is to have a hot bottom burning. It is thought that more 
%massive stars undergo hot-bottom burning (HBB) that effectively converts
%dredged-up carbon into nitrogen (Herwig 2005).  CRL 618 is also nitrogen rich
%(Calvet and Primbert 1983).
%BUt HBB also destroys 15N
%CRL 618 is rich not only in $^{14}$N but also in $^{15}$N.

%One more possibility is to have deep mixing inside the star, as suggested by
%\citet{Palmerini2011}.

% CNO cycle in a hot bottom burning \citet{Frost1998}. 

%Note that, however, this trend seems to be in opposite
%to the conclusion of \citet{Milam2009}

% In Wannier 1991
% 8 carbon-rich envelope  
% AGB stars: IRC+10216,  CRL3068, CIT 6, IRC+40540
% PPNs: CRL618 CRL2688
% RGB: VCyn

% However, as mentioned above, the
%$^{12}$C/$^{13}$C ratio is not a constant, and it decreases from $\gtrsim40$
%in the extended halo to $\sim$ 9 in the dense core.  Thus, the
%$^{14}$N/$^{15}$N ratio could also decrease from $>400$ in the extended halo
%to $>90$ in the dense core.  As a result, our value in the dense core is
%consistent with their measurement considering the decrease in
%$^{12}$C/$^{13}$C ratio.  

\subsubsection{Nitrogen}

HC$^{15}$N is clearly detected here in the dense core in J=4-3 line.  It was
also detected in J=1-0, 2-1, and 3-2 lines by \citet{Pardo2007b} (see their
electronic version of Fig.  5), although it was not listed in their Table 2. 
The isotopic ratio of $^{14}$N/$^{15}$N in the dense core can thus be
derived by dividing the abundance of HCN by that of HC$^{15}$N and is found
to be $\sim$ 130$\pm40$.  As discussed later, this value is low
compared to those found in AGB stars \citep{Wannier1991}.  It could be due
to a possible underestimate of the HCN abundance in our model, considering
that \citet{Thorwirth2003} has detected direct l-doubling transition lines
of HCN (for which the line strengths are very low).  We also derive an
independent estimate of the $^{14}$N/$^{15}$N ratio, $\sim$ 160$\pm$40, by
multiplying the [H$^{13}$CN]/[HC$^{15}$N] ratio (16$\pm$4) by the
$^{12}$C/$^{13}$C ratio ($\sim$10, derived earlier from HC3N and its
isotopologues).
 Therefore, the mean ratio of $^{14}$N/$^{15}$N can be 150$\pm50$.  With
this ratio, our model predicts a weak HCCC$^{15}$N (0000) J=39-38 line at
344385.3481 MHz (see the green spectrum in Fig.  \ref{fig:fitspec}), roughly
consistent with the line emission feature tentatively detected there. 
However, such a feature could be alternatively identified with weak emission
from the \HCC13CN{} (0100)/(0003) J=38-37 (f component) transition (see the
red spectrum in Fig.  \ref{fig:fitspec}).  Further work is needed to check
this possible detection of the HCCC$^{15}$N (0000) line.

%Further observations
%with higher sensitivities
%  however,
% unfortunately no other species can be used to confirm our value.  

In \citet{Pardo2007b}, CN and C$_3$N were not detected in the dense core (or
their SEE), and neither were their $^{15}$N-isotopologues.  Although HNC and
HC$_3$N were also detected in the dense core, their $^{15}$N-isotopologues
were not detected likely because of not enough sensitivity in their
observations as discussed below.   Since HNC lines are likely optically
thick, we use HC$^{15}$N lines in their observations to estimate the
expected peak intensities for the H$^{15}$NC lines. Assuming the abundance
ratio of [HC$^{15}$N]/[H$^{15}$NC]=[HCN]/[HNC]= 10 \cite[as found in the SEE
by][]{Pardo2007b} and that the lines are optically thin, then the H$^{15}$NC
lines are expected to have a peak of $\sim$ 1 mK, 10 mK and 12 mK,
respectively for J=1-0 (89 GHz), J=2-1 (178 GHz), and J=3-2 (267 GHz) lines. 
In their observations, the noise was $\sim$ 4 mK at 3mm (100 GHz), 8 mK at
2mm (150 GHz), 11 mK from 204 to 240 GHz, and 14 mK above 240 GHz, and thus
those lines were lost in the noise.  As for HCCC$^{15}$N, here we only check
if its (0000) lines can be detected because its higher vibrational lines are
much weaker.  In \citet{Pardo2007b}, the HC$_3$N (0000) lines have a peak of
0.4-0.6 K for those below 100 GHz and $\sim$ 1 K for those above 140 GHz. 
Assuming $^{14}$N/$^{15}$N =150 (as derived above) and the lines are
optically thin, then HCCC$^{15}$N (0000) lines are expected to have a peak
of $\sim$ 3 mK below 100 GHz and 7 mK above 140 GHz, and thus
were also lost in the noise -- more sensitive observations are needed.

%  However, since HC$_3$N (0000) lines are likely
%optically thick, further observations at higher sensitivity are needed to
%check our conclusion.

%Noise = 4 mK at 3mm (100GHz), 8mk at 2mm (150 GHz) and 11mk from 204 to 240 GHz and 14mk above 240 GHz

%On the other hand, the isotopic ratio of $^{14}$N/$^{15}$N is $\sim$
%200$\pm60$ comparing the abundances of \nHC3N and its isotopologues, and is
%$\sim$ 117$\pm40$ comparing the abundances of HCN and its isotopologues. 
%Thus, the mean value of this ratio is $\sim$ 160$\pm90$.  

Previously, using
single-dish observations in \nHC3N lines at lower frequencies,
\citet{Wannier1991} derived a
$^{15}$N$\times$$^{12}$C/$^{14}$N$\times$$^{13}$C ratio of $<0.12$. 
Assuming a $^{12}$C/$^{13}$C ratio of $\gtrsim$ 30, they estimated a
$^{14}$N/$^{15}$N ratio of $>250$.  However, since the $^{12}$C/$^{13}$C
ratio in CRL618 decreases from $\gtrsim$ 40 in the extended halo to $\sim$ 9
in the dense core, the Wannier et al. lower limit on the $^{14}$N/$^{15}$N
ratio is $>$ 75 in the dense core, consistent with our derived value. 
\citet{Wannier1991} found the $^{14}$N/$^{15}$N ratio to be $>$ 500 for
their small sample of a few carbon-rich AGB stars and one PPN;
\citet{Zhang2009} confirm this for AFGL 3068, finding
$^{14}$N/$^{15}$N$=$1099.  CRL 618 appears to be different from these with
its lower $^{14}$N/$^{15}$N ratio.  However, these studies use single-dish
observations and their derived ratios are likely characteristic of the
extended circumstellar envelopes in these objects, and not their dense
central regions.  High-resolution studies of these objects, like the one
presented here, should be carried out to probe the latter.

%\citet{Wannier1991} also found the
%$^{14}$N/$^{15}$N ratio to be $>$ 500 for a few carbon-rich AGB stars and
%another PPN.  However, since they used the single-dish observations, their
%derived ratios could be more for the extended halo.  Further observations at
%higher resolution are needed to determine the ratio near the stars for an
%appropriate comparison.

Our value of $^{14}$N/$^{15}$N is smaller than the solar value, which is
$\sim$ 272.  In current models of nucleosynthesis in the evolved stars,
however, the CNO cycle is a cold CNO cycle that destroys $^{15}$N, resulting
in a $^{14}$N/$^{15}$N ratio always $\gtrsim$ 2000 \cite[see,
e.g.,][]{Palmerini2011}.  The only known way to produce abundant $^{15}$N is
through a hot CNO cycle as in novae \citep{Wiescher2010}.  However, it is
unclear if the hot CNO cycle can really take place in the AGB stars at the
end of the AGB phase.  Also, chemical fractionation that can enhance
the abundance of the isotopically-substituted species, is unlikely to be an
effect at the high temperatures of the dense core, because it is important
only at low temperatures \citep{Terzieva2000}.

\section{Conclusions}

With the SMA, we have mapped the central region of CRL 618 in continuum and
molecular lines at 350 GHz at \arcsa{0}{3} to \arcsa{0}{5} resolution.  Most
of the emission detected in our observations arises within a radius of
$\sim$ 630 AU (\arcsa{0}{7}) from the source.  The main conclusions are the
following: \begin{itemize}

\item In the continuum, there are two components, (1) a compact emission at
the center tracing the dense inner part of the H II region previously
detected in 23 GHz continuum and it may trace a fast ionized wind at the
base, and (2) an extended emission tracing the thermal dust emission from
the dense core around the H II region.  The dense core seems to have a
density enhancement in the equatorial plane that can confine the H II region
into a bipolar morphology.  The dust emissivity index is estimated to be
$0.8\pm0.2$, suggesting that the dust grains in the dense core may have
grown to mm size.

\item The dense core is also detected in \nHC3N, HCN, and their
isotopologues, with higher excitation lines tracing closer to the source. 
It is also detected in \C2H3CN{} toward the innermost part.  The dense core
detected here is the inner part of that seen before in a lower excitation
line of \nHC3N{}, and it could also have a density enhancement in the
equatorial plane.  The dense core is probably also excavated by the outflow
lobes.

\end{itemize}

We have fitted a simple radiative transfer model to our observations in
order to derive the kinematics, physical conditions, and the chemical
abundances in the dense core.  In this model, the H II region is ellipsoidal
at the center.  The dense core is spherical with the inner boundary set by
the outer boundary of the H II region and the outer radius set to
\arcsa{1}{2}.  It is dusty and molecular, producing both the thermal dust
emission and molecular emission.  Two outflow cavities are also included. 
This simple model can roughly fit the observations.   The model results are the following:

\begin{itemize}

\item The dense core is expanding, with the velocity increasing roughly
linearly from $\sim$ 3 \vkm{} in the innermost part to $\sim$ 16 \vkm{} at
630 AU.  The mass-loss rate in the dense core is extremely high with a value
of $\sim 1.15\times 10^{-3}$ \smpy{}.  The dense core has a mass of $\sim$ 0.47
$\solarM{}$ and a dynamical age of $\sim$ 400 yrs.  It could result from a
recent enhanced heavy mass-loss episode that ends the AGB phase.

\item The isotopic ratios of $^{12}$C/$^{13}$C and $^{14}$N/$^{15}$N are
$\sim$ 9$\pm4$ and $150\pm50$, respectively, both smaller than the solar
values.  The $^{12}$C/$^{13}$C ratio is also much smaller than that found in
the extended halo, indicating that this ratio decreases toward the center,
as argued before.  It is not clear if current models of nucleosynthesis in
evolved stars can produce our isotopic ratios in a C-rich star like that in
CRL 618.
 
\end{itemize}

\acknowledgements
We thank the anonymous referee for the valuable and insightful comments.
We thank the SMA staff for their efforts in running and maintaining the
array. C.-F. Lee and C.-H. Yang acknowledge grants from the National
Science Council of Taiwan (NSC 99-2112-M-001-007-MY2 and NSC 101-2119-M-001-002-MY3).
CSC has been partially supported by the Spanish MICINN/MINECO through grants
AYA2009-07304, AYA2012-32032, and CONSOLIDER INGENIO 2010 for the team
``Molecular Astrophysics: The Herschel and ALMA Era -- ASTROMOL'' (ref.:
CDS2009-00038).

% C3H3CN has a log(intensity) greater (-2.6)
% Note that there is no C3H3CN from -2.6 to -3
% except for 342123.5509, which happened to have 2 lines each with -2.9979.
% the total strenths of the two lines are -2.7
\begin{deluxetable}{llr}
\tablecolumns{3}
\tabletypesize{\normalsize}
\tablecaption{Line Detections
\label{tab:linedetection}}
\tablewidth{0pt}
\tablehead{
\colhead{Frequency} & \colhead{Species and} & \colhead{Rotational} \\
\colhead{(MHz)} &\colhead{Vibrational State$^a$} & \colhead{Transition J(QN)}      
}
\startdata
 342123.5509 & \C2H3CN    \vn{} &      J=36(12,24)-35(12,23) \\  
 342123.5509 & \C2H3CN    \vn{} &       J=36(12,25)-35(12,24)  \\
% 342203.3300 & \CH2CN  ? &        J=17-16 \\ 
 342204.9747 & H$^{13}$CC$^{13}$CN (0000) &    J=39-38  \\
% 342208.5700 & \CH2CN ?  &        J=17-16  \\
 342317.5544 & \C2H3CN    & \vn{}       J=36(5,32)-35(5,31) \\  
 342375.5639 & \C2H3CN    & \vn{}      J=36(5,31)-35(5,30)  \\
% 342467.9204 & H$^{13}$C$^{13}$CCN (0000) &    J=39-38 (No detection) \\
 342585.4708 & \C2H3CN    & \vn{}       J=36(4,33)-35(4,32) \\  
 342666.3834 & HC$^{13}$C$^{13}$CN (0000) &   J=38-37  \\
 342882.8503 & CS        &      J= 7- 6  \\
 343446.5355 & \C2H3CN    & \vn{}       J=36(4,32)-35(4,31) \\  
 343737.3998 & \H13CCCN   (0000) &      J=39-38  \\
 344142.9833 & \HC13CCN{} (0000) &     J=38-37  \\
 344176.1674 & \HCC13CN   (0000) &     J=38-37  \\
 344198.9411 & \HCC13CN   (0100)/(0003)?  &  J=38-37 \\  %suggested by referee 1e line
 344200.1089 & HC$^{15}$N     & \vn{}      J=4-3 \\ 
 344302.9375 & \H13CCCN   (0010) &     J=39-38  \\
 344385.3481 & HCCC$^{15}$N (0000)? &    J=39-38  \\  % challenged by the referee 1f line
 344390.9656 & \HCC13CN  (0100)/(0003)? &    J=38-37 \\ %suggested by referee 1f line
 344565.1094 & \H13CCCN   (0010) &     J=39-38  \\
 344590.0711 & \H13CCCN   (0001) &    J=39-38  \\
 344694.4713 & \HC13CCN   (0010) &    J=38-37  \\
 344719.4221 & \HCC13CN   (0010) &    J=38-37  \\
 344955.5335 & \HC13CCN   (0001) &    J=38-37  \\
 344967.3969 & \HC13CCN   (0010) &    J=38-37  \\
 344994.6742 & \HCC13CN   (0010) &    J=38-37  \\
 345003.2278 & \HCC13CN   (0001) &    J=38-37  \\
% 345050.0900 & ? \CH2CN     &          J=17-16  \\
% 345066.0000 & ? \CH2CN     &          J=17-16  \\
 345068.6323 & \H13CCCN   (0001) &    J=39-38  \\
 345122.5871 & \nHC3N     (1000) &    J=38-37  \\
 345238.7103 & H$^{13}$CN  \v{2}{1} &     J= 4- 3  \\
 345339.7694 & H$^{13}$CN  \vn{}  &    J= 4- 3  \\
% 345342.6000 & HNCCC     &          J=37-36  \\
% 345343.0100 & HNCCC     &      J=37-36  \\
% 345343.0200 & HNCCC     &      J=37-36  \\
% 345343.0200 & HNCCC     &      J=37-36  \\
% 345343.0200 & HNCCC     &      J=37-36  \\
% 345343.4200 & HNCCC     &      J=37-36  \\
 345451.7567 & \HC13CCN   (0001) &     J=38-37 \\  
 345495.8736 & \HCC13CN   (0001) &    J=38-37 \\
 345609.0100 & \nHC3N     (0000) &     J=38-37 \\  
 345632.1266 & \nHC3N     (0100)/(0001) &   J=38-37 \\  
 345795.9899 & CO         \vn{}  &          J= 3- 2 \\  
 345797.0584 & \H13CCCN   (0002) &    J=39-38  \\
 345824.6653 & \nHC3N     (0100)/(0001) &   J=38-37 \\  
 345862.8068 & \nHC3N     (1002)/(0200) & J=38-37 \\
 345917.7699 & \H13CCCN   (0002) &    J=39-38  \\
 346010.5711 & \nHC3N     (1001) &   J=38-37 \\  
 346041.2173 & \H13CCCN   (0002) &   J=39-38  \\
 354127.5000 & U         &                  \\
 354143.5000 & U         &                  \\
 354197.5820 & \nHC3N     (1000) &        J=39-38 \\  
 354460.4346 & HCN       \v{2}{1} &         J= 4- 3 \\ 
 354505.4773 & HCN        \vn{}  &        J= 4- 3 \\
 354535.7250 & \HC13CCN   (0001) &     J=39-38  \\
 354580.9970 & \HCC13CN   (0001) &    J=39-38  \\
 354650.5747 & \H13CCCN   (0002) &    J=40-39  \\ 
% 354669.7700 & HNCCC     &       J=38-37  \\
% 354670.1800 & HNCCC     &      J=38-37  \\
% 354670.1900 & HNCCC     &      J=38-37  \\
% 354670.1900 & HNCCC     &      J=38-37  \\
% 354670.1900 & HNCCC     &      J=38-37  \\
% 354670.6000 & HNCCC     &      J=38-37  \\
 354697.4631 & \nHC3N     (0000) & J=39-38  \\
 354721.1070 & \nHC3N     (0100)/(0001) &   J=39-38   \\
 354780.4494 & \H13CCCN   (0002) &    J=40-39  \\
 354913.0924 & \H13CCCN   (0002) &    J=40-39  \\
 354918.6770 & \nHC3N     (0100)/(0001) &   J=39-38 \\  
 354957.7936 & \nHC3N     (1002)/(0200) & J=39-38  \\
 355108.4000 & \nHC3N     (1001) &    J=39-38 \\ 
 355213.8330 & \HC13CCN   (0002) &    J=39-38  \\
 355277.5940 & \nHC3N     (0010) &        J=39-38 \\  
 355281.4277 & \HCC13CN   (0002) &    J=39-38  \\
 355317.9566 & \C2H3CN     \vn{} &      J=38(2,37)-37(2,36)  \\
 355365.4210 & \HC13CCN   (0002) &     J=39-38  \\
 355424.7668 & \HCC13CN   (0002) &    J=39-38  \\
 355463.0140 & \nHC3N     (0110) &     J=39-38 \\  
 355520.9620 & \HC13CCN   (0002) &    J=39-38  \\
 355533.6650 & \nHC3N     (0110) &    J=39-38 \\  
 355544.9810 & \nHC3N     (0110) &     J=39-38 \\
 355556.5100 & \nHC3N     (0010) &       J=39-38 \\
 355566.2540 & \nHC3N     (0001) &        J=39-38 \\
 355572.7017 & \HCC13CN   (0002) &    J=39-38  \\
 355629.3710 & \nHC3N     (1001) &    J=39-38 \\ 
 355678.1620 & \nHC3N     (0101) &    J=39-38  \\
 355729.5000 & \nHC3N    (0110) &    J=39-38  \\  %suggested by referee
 %355756.0000 & U         &                     \\
 355751.3900 & \nHC3N     (0201) &    J=39-38  \\ % suggested by referee
 355837.4410 & \nHC3N     (0020) &        J=39-38  \\
 355910.9550 & \nHC3N     (0101) &    J=39-38  \\
 355986.3120 & \nHC3N     (0101) &    J=39-38  \\
 356072.4450 & \nHC3N     (0001) &       J=39-38  \\
 356125.6170 & \nHC3N     (0020) &       J=39-38  \\
 356131.7500 & \nHC3N     (0020) &       J=39-38  \\
 356135.4600 & HCN      \v{2}{2} &         J= 4- 3  \\
 356158.0740 & \nHC3N     (0101) &    J=39-38  \\
 356162.7700 & HCN      \v{2}{2} &         J= 4- 3  \\
 356247.4196 & \C2H3CN    \vn{}  &     J=39(1,39)-38(1,38)  \\
 356255.5682 & HCN      \v{2}{1} &        J= 4- 3 \\  
 356301.1780 & HCN      \v{2}{2} &        J= 4- 3  \\
 356349.7990 & \nHC3N     (0011) &    J=39-38  \\
 356421.7169 & \C2H3CN     \vn{} &       J=39(0,39)-38(0,38)  \\
 356449.5064 & \nHC3N    (1002)/(0200) & J=39-38 \\ 
 356461.9430 & \nHC3N     (0011) &    J=39-38 \\
% 356503.0000 & U         &                \\
% 356515.3000 & U         &               \\
 356511.7310 & \nHC3N     (0111)?  &   J=39-38            \\ %suggested by the referee
 356570.8112 & \nHC3N    (1002)/(0200) & J=39-38  \\
 356632.0040 & \nHC3N     (0011) &    J=39-38  \\
 356647.4090 & \nHC3N    (0111)?  &   J=39-38            \\ %added in response to other line suggested by the referee
 356650.3550 & \nHC3N    (0111)?  &   J=39-38            \\ %added in response to other line suggested by the referee
 356716.0590 & \nHC3N    (1002)/(0200) & J=39-38  \\
 356734.2230 & HCO$^+$     \vn{} &        J= 4- 3  \\
 356742.1080 & \nHC3N     (0011) &     J=39-38  \\
 356798.5018 & \nHC3N     (0002) &       J=39-38  \\
 356832.0039 & \C2H3CN     \vn{} &      J=37(3,34)-36(3,33)  \\
 356937.1366 & \nHC3N     (0002) &        J=39-38 \\ 
 357079.4470 & \nHC3N     (0002) &       J=39-38  \\
 357129.0100 & \nHC3N     (0102) &    J=39-38 \\ 
 357202.0855 & \C2H3CN    \vn{}  &      J=37(2,35)-36(2,34)  \\
 357240.0820 & \nHC3N     (0102) &    J=39-38 \\ 
 357254.0400 & \nHC3N     (0012) &   J=39-38  \\
 357443.2270 & \nHC3N     (0102) &    J=39-38 \\ 
 357497.4370 & \nHC3N    (0100)/(0001) &  J=39-38  \\
 357530.2450 & \nHC3N     (0012) &   J=39-38 \\ 
% 357559.0000 & ? U         &                   \\
% 357566.9800 & HCCNC     &      J=36-35   \\ 
% 357567.3300 & HCCNC     &      J=36-35  \\
% 357567.3400 & HCCNC     &      J=36-35  \\
% 357567.3400 & HCCNC     &      J=36-35  \\
% 357567.3400 & HCCNC     &      J=36-35  \\
 357567.3400 &  HCCNC     &      J=36-35  \\
% 357567.6900 & HCCNC     &      J=36-35  \\
% 357574.0000 & ? U        &               \\
 357740.0520 & \nHC3N     (0012) &    J=39-38  \\
 357759.7070 & \nHC3N     (0012) &   J=39-38  \\
 357871.2740 & \nHC3N     (0012) &   J=39-38  \\
 357885.9650 & \nHC3N     (0012) &   J=39-38  \\
 357920.9088 & \C2H3CN     \vn{} &      J=38(1,37)-37(1,36)  \\
\enddata

\tablenotetext{a}{For \nHC3N and its isotopologues, the vibrational states
are $(v_4v_5v_6v_7)$.  A ``/" in the vibrational state indicates a
Fermi resonance transition.  The rest frequencies of the \nHC3N{} lines are
mostly obtained from \citet{Mbosei2000}.  A ``?" indicates a possible
detection.}

%\tablenotetext{all}{D: dominated, *:blended, e: at the edge, S: separatable, w:very weak}
\end{deluxetable}

\begin{deluxetable}{lrrcccc}
\tablecolumns{6}
\tabletypesize{\normalsize}
\tablecaption{Line Information and Measurements of \nHC3N, H$^{13}$CCCN, HC$^{13}$CCN, and HCC$^{13}$CN
\label{tab:measure}}
\tablewidth{0pt}
\tablehead{
\colhead{Molecule and} & \colhead{J} & \colhead{Frequency} 
& \colhead{E$_u$} & \colhead{$T_B^p$} & $\triangle v$ & Diameter \\
\colhead{Vibrational State} & \colhead{Transition}  & \colhead{(MHz)}       & 
\colhead{(K)}   & \colhead{(K)} & \colhead{(\vkm{})} & \colhead{(")}
}
\startdata
\nHC3N{}&&&&& & \\
\hline
(1000)        & J=38-37 &   345122.5871 &   1590.042 &  25.0 &  8.0 & 0.59\\ % 178 HC3Nv4=1        J=38-37  6 to 8.3, absorp at -7.6
(0000)i        & J=38-37 &   345609.0100 &    323.493 &  77.0    & 17.0     & 1.12 \\ %Separable
%(0100)/(0001) & J=38-37 &   345632.1266 &   1277.523 & *18.0 &  7.0 & -- \\ % 119 HC3Nv5=1_v7=1   J=38-37  f detected, blended, extended?
(1001)        & J=38-37 &   346010.5711 &   1907.628 &  7.0 &  6.5 & --  \\ %  41 HC3Nv4=v7=1     J=38-37  4 to 7, weak
(1000)        & J=39-38 &   354197.5820 &   1607.041 &  22.0 &  8.0 & 0.57\\ % 191 HC3Nv4=1        J=39-38  6 to 6.5, compact, slightly per. to jet
(0000)i        & J=39-38 &   354697.4631 &    340.516 &  77.0 & 17.5 & 1.09 \\ %1303 separable HC3Nv=0         J=39-38  t least -38 to 38 , torus and shell
%(0100)/(0001) & J=39-38 &   354721.1070 &   1294.547 &  *20.0 &  7.0 & -- \\ % 126 HC3Nv5=1_v7=1   J=39-38  eak and lost in HC3Nv=0
(0100)/(0001) & J=39-38 &   354918.6770 &   1294.737 &  40.0 & 10.0 & 0.60 \\ % 435 dominated HC3Nv5=1_v7=1   J=39-38  9 to 13, strong, red side is H13CCCNv7=2
%(1002)/(0200) & J=39-38 &   354957.7936 &   2245.575 &  no det & -- & \\ %  22 HC3Nv4=1v7=2_v5 J=39-38  o clear detection.
(1001)        & J=39-38 &   355108.4000 &   1924.670 &  7.0 &  7.0 & 0.50\\ %  42 HC3Nv4=v7=1     J=39-38  4 to 3, weak or no detection
(0010)        & J=39-38 &   355277.5940 &   1058.752 &  50.0 & 10.5 & 0.65 \\ % 614 dominated HC3Nv6=1        J=39-38  10 to 10 compact
%(0110)        & J=39-38 &   355463.0140 &   1994.135 &  no det &  -- & \\ %  40 HC3Nv5=v6=1     J=39-38  5 to 4, weak toroidal
%(0110)        & J=39-38 &   355533.6650 &   1994.135 &  *8.0 &  5.0 & -- \\ %in the tail
%(1001)        & J=39-38 &   355629.3710 &   1925.175 &  *9.0 &  6.0 & -- \\ %  70 HC3Nv4=v7=1     J=39-38  3 to 5. compact
(0101)        & J=39-38 &   355678.1620 &   1597.035 &  21.0 &  8.0 & 0.53 \\ % 170 HC3Nv5=v7=1     J=39-38  5.8 to 6.5, compact
(0020)        & J=39-38 &   355837.4410 &   1766.135 &  14.0 &  7.0 & 0.61 \\ % 104 HC3Nv6=2        J=39-38  5.7 to 5.2, compact
(0101)        & J=39-38 &   355910.9550 &   1597.035 &  17.0 &  8.0 & 0.54 \\ % 136 HC3Nv5=v7=1     J=39-38  6.8 to 6.8, compact perp to jet
(0101)        & J=39-38 &   355986.3120 &   1597.035 &  19.0 &  8.5 & 0.54 \\ % 171 HC3Nv5=v7=1     J=39-38  5 to 6, rotating compact?
(0001)i        & J=39-38 &   356072.4450 &    662.686 &  73.0 & 13.0 & 0.86 \\ % 992 HC3Nv7=1        J=39-38  8 to 14, blueside went out the window, extended perp jet. torus.
%(0020)        & J=39-38 &   356125.6170 &   1766.135 &  *18.5 &  6.0 & --\\ % 118 HC3Nv6=2        J=39-38  lended merge with HC3Nv6=2
%(0020)        & J=39-38 &   356131.7500 &   1766.135 &  *18.5 &  6.0 & --\\ % 118 HC3Nv6=2        J=39-38  lended -7 to 9.5, compact
%(0101)        & J=39-38 &   356158.0740 &   1597.035 &  *24.5 &  8.0 & --\\ % 208 HC3Nv5=v7=1     J=39-38  8 to 8, Rotating compact
(0011)        & J=39-38 &   356349.7990 &   1378.505 &  38.0 &  8.5 & 0.53 \\ % 289 HC3Nv6=v7=1     J=39-38  7 to 8.5, compact
%(1002)/(0200) & J=39-38 &   356449.5064 &   2242.684 &  u*6 &  5 & -- \\ %  14 HC3Nv4=1v7=2_v5 J=39-38  no detection.
(0011)        & J=39-38 &   356461.9430 &   1379.687 &  31.0 &  9.0 & 0.60 \\ % 278 dominated HC3Nv6=v7=1     J=39-38  8 to 9.7 strong elongated perp. to jet axis, faint shell
(1002)/(0200) & J=39-38 &   356570.8112 &   2246.276 &  3.5 &  8.0 & -- \\ %  29 HC3Nv4=1v7=2_v5 J=39-38  6.7 to 5.6, elongated perp to the jet, but why no detection for the same vibrational a
(1002)/(0200) & J=39-38 &   356570.9650 &   2246.276 &  3.5 &  8.0 & -- \\ %  29 HC3Nv4=1v7=2_v5 J=39-38  6.7 to 5.6, elongated perp to the jet, but why no detection for the same vibrational a
(0011)        & J=39-38 &   356632.0040 &   1379.991 & 33.0 &  9.5 & 0.54 \\ % 276 HC3Nv6=v7=1     J=39-38  7 to 10.6, compact, peak at 0.05" to North
%(0002)        & J=39-38 &   356798.5018 &    984.295 & *44.0 & 10.5 & -- \\ % 424 HC3Nv7=2        J=39-38  9 to 8.5 elongated perp to the jet axis, HCO+ shell
%(0002)        & J=39-38 &   356937.1366 &    987.636 & e54.0 & 10.5 & --\\ % 479 HC3Nv7=2        J=39-38  10 to 13.4, extended array only
(0002)i        & J=39-38 &   357079.4470 &    987.708 &  51.0 & 11.0 & 0.65 \\ % 503 HC3Nv7=2        J=39-38  8 to 15, elongated perp to the jet axis
%(0102)        & J=39-38 &   357240.0820 &   1919.319 &  w5 &  6.5 & -- \\ %  29 HC3Nv5=1v7=2    J=39-38  eak
(0012)        & J=39-38 &   357254.0400 &   1683.361 &  15.0 &  6.0 & -- \\ %  86 HC3Nv6=1v7=2    J=39-38  5 to 3, 0.1 to the north
%(0102)        & J=39-38 &   357443.2270 &   1919.319 &  ?14.5 &  3.0 & -- \\ %  46 HC3Nv5=1v7=2    J=39-38  4 to 4 ? compact peak 0.1 to the east
(0100)/(0001) & J=39-38 &   357497.4370 &   1298.070 &  38.0 &  8.5 & 0.58 \\ % 294 HC3Nv5=1_v7=1   J=39-38  8 to 9.7, elongated perp the jet
(0012)        & J=39-38 &   357530.2450 &   1683.361 &  15.5 &  9.0 & 0.48 \\ % 148 HC3Nv6=1v7=2    J=39-38  6.4 to 8.6, compact peak near the center
(0012)        & J=39-38 &   357740.0520 &   1683.361 &  13.0 &  9.0 & 0.60 \\ % 129 HC3Nv6=1v7=2    J=39-38  -7 to 11, compact 0.1" north, elongated perp to jet
(0012)        & J=39-38 &   357759.7070 &   1683.361 &  13.5 &  8.0 & 0.60 \\ % 114 HC3Nv6=1v7=2    J=39-38  7 to 7, compact peak 0.07 to the SE
(0012)        & J=39-38 &   357871.2740 &   1683.361 &  12.5 &  8.5 & 0.54 \\ % 119 HC3Nv6=1v7=2    J=39-38  -6 to 8, compact 0.1 South
(0012)        & J=39-38 &   357885.9650 &   1683.361 &  15.0 &  6.5 & --  \\ % 100 HC3Nv6=1v7=2    J=39-38  5.6 to 4, compact 0.05 to NE
\hline
H$^{13}$CCCN &&&&& & \\
\hline
(0000)        & J=39-38 &   343737.3998 &    329.992 & 55.0 & 11.0 & 0.76 \\ % 574 H13CCCNv=0      J=39-38  trong, modeled by Chun-Hui
%(0010)        & J=39-38 &   344302.9375 &   1047.651 &  w8.0 &  6.0 & -- \\ %  45 H13CCCNv6=1     J=39-38  5 to 5 km, weak
(0010)        & J=39-38 &   344565.1094 &   1047.903 &  8.5 &  7.5 & 0.52 \\ %  52 H13CCCNv6=1     J=39-38  4.6to 6.9
(0001)        & J=39-38 &   344590.0711 &    648.992 & 28.0 &  9.0 & 0.60 \\ % 229 H13CCCNv7=1     J=39-38  6 to 10.4
(0001)        & J=39-38 &   345068.6323 &    649.453 & 28.0 &  8.0 & 0.61 \\ % 204 H13CCCNv7=1     J=39-38  6 to 9.4, absorp at blue
%(0002)        & J=39-38 &   345917.7699 &    971.458 & *8.5 &  6.5 & -- \\ %  48 H13CCCNv7=2     J=39-38  5 to 6 Only the part around the sourceo
(0002)        & J=39-38 &   346041.2173 &    971.520 &  8.5 &  7.5 & 0.58\\ %  55 H13CCCNv7=2     J=39-38  4 to 7, weak
%(0002)        & J=40-39 &   354650.5747 &    985.020 & *6.5 &  7.0 & -- \\ %  51 H13CCCNv7=2     J=40-39  6 to 6, highly contaminated by HC3Nv=0 shell
%(0002)        & J=40-39 &   354780.4494 &    988.485 & *9.0 &  8.0 & -- \\ %  95 H13CCCNv7=2     J=40-39  6 to 7, contaminated by HC3Nv=0 shell
\hline
HC$^{13}$CCN &&&&& & \\
\hline
(0000)        & J=38-37 &   344142.9833 &    322.121 &  54.0 & 11.0 & 0.74 \\ % 542 HC13CCNv=0      J=38-37  trong
(0010)i        & J=38-37 &   344694.4713 &   1030.321 & 5.0 &  7.5 & --\\ %  39 HC13CCNv6=1     J=38-37  4.9 to 6.3
(0001)        & J=38-37 &   344955.5335 &    638.768 & 23.0 & 10.0 & 0.60 \\ % 234 HC13CCNv7=1     J=38-37  7 to -9
(0010)i        & J=38-37 &   344967.3969 &   1030.577 &  6.0 &  5.0 & --\\ %  26 HC13CCNv6=1     J=38-37  lended slightly at lower freq
(0001)        & J=38-37 &   345451.7567 &    639.234 & 24.0 &  8.5 & 0.57 \\ % 199 HC13CCNv7=1     J=38-37  6.5 to 7.7
(0002)        & J=39-38 &   355213.8330 &    972.521 & 13.0 &  8.5 & 0.58 \\ % 108 HC13CCNv7=2     J=39-38  6.7 to 8.3, compact
(0002)        & J=39-38 &   355365.4210 &    975.606 &  8.0 &  9.0 & 0.57 \\ %  76 HC13CCNv7=2     J=39-38  7 to 7, compact with weak shells
(0002)        & J=39-38 &   355520.9620 &    975.684 & 10.0 &  7.5 & 0.46 \\ %  93 HC13CCNv7=2     J=39-38  5.4 to 5.5 compact, could be contanimated by HC3Nv5=v6=1 on the blue
\hline
HCC$^{13}$CN &&&&& & \\
\hline
(0000)        & J=38-37 &   344176.1674 &    322.152 &  51.0 & 11.5 & 0.74 \\ % 550 HCC13CNv=0      J=38-37  lended slightly with HC15Nv=0
(0010)        & J=38-37 &   344719.4221 &   1025.870 &  8.0 &  7.0 & -- \\ %  47 HCC13CNv6=1     J=38-37  
%(0010)        & J=38-37 &   344994.6742 &   1026.128 &  *6.0 &  6.0 & -- \\ %  38 HCC13CNv6=1     J=38-37  lended! Can't map
%(0001)        & J=38-37 &   345003.2278 &    641.263 &  *26.0 &  9.0 & --  \\ % 220 HCC13CNv7=1     J=38-37  lended slight at lower freq
(0001)        & J=38-37 &   345495.8736 &    641.726 &  25.0 &  8.5 & 0.58 \\ % 217 HCC13CNv7=1     J=38-37  6.2 to 7.9
%(0001)        & J=39-38 &   354580.9970 &    658.743 &  *18.0 &  8.5 & -- \\ % 151 HCC13CNv7=1     J=39-38  6 to 9, highly contaminated by HCNv=0 shell
(0002)        & J=39-38 &   355424.7668 &    979.687 &  8.5 &  6.0 & --\\ %  57 HCC13CNv7=2     J=39-38  4 to 5
\enddata
%\tablenotetext{a}{The radius is the half of the emission size at half
%maximum. It can be estimated for the uncontaminated lines and those lines
%that are strong enough and those lines that are resolvable with current
%resolution.}
\tablenotetext{i}{The data will not be used for the population diagram fitting.}
%\tablenotetext{1}{Used for moment map}
%\tablenotetext{2}{Too weak to determine the peak and width}
%\tablenotetext{all}{D: dominated, *:blended, e: at the edge, S: separable, w:very weak}
%\tablenotetext{d}{Observational result from \citet{LMT2001}}
%\tablenotetext{\dagger}{Synthesized beam size in the channel maps.}
%\tablenotetext{e}{Noise level in the channel maps with a velocity resolution of 1 \vkm{}.}
%\tablenotetext{f}{Noise level in the channel maps with a velocity resolution of 0.1 \vkm{}.}
\end{deluxetable}

\begin{deluxetable}{lr}
\tablecolumns{6}
\tabletypesize{\normalsize}
\tablecaption{Model Parameters
\label{tab:model}}
\tablewidth{0pt}
\tablehead{
\colhead{Species} & \colhead{Abundance}
%\colhead{(K)}   & \colhead{(K)} & \colhead{(\vkm{})} & \colhead{(")}
}
\startdata
\nHC3N         & $2\pm0.4 \times 10^{-7}$  \\
\begin{math}
\left. 
\hspace{-0.21cm}
\begin{array}{l}
   \textrm{\H13CCCN} \\
   \textrm{\HC13CCN} \\
   \textrm{\HCC13CN} 
\end{array}
\right\}   
\end{math}
 & $2\pm0.4 \times 10^{-8}$ \\
%\H13CCCN       & $2 \times 10^{-8}$  \\ 
%\HC13CCN       & $2 \times 10^{-8}$  \\ 
%\HCC13CN       & $2 \times 10^{-8}$  \\ 
%HCCC$^{15}$N       & $1\pm0.2 \times 10^{-9}$  \\
\begin{math}
\left. 
\hspace{-0.21cm}
\begin{array}{l}
   \textrm{H$^{13}$C$^{13}$CCN} \\
   \textrm{HC$^{13}$C$^{13}$CN} \\
   \textrm{HCC$^{13}$C$^{13}$N} 
\end{array}
\right\}   
\end{math}
 & $2\pm0.4 \times 10^{-9}$ \\
%H$^{13}$C$^{13}$CCN       & $2 \times 10^{-9}$ \\
%HC$^{13}$C$^{13}$CN       & $2 \times 10^{-9}$ \\
%HCC$^{13}$C$^{13}$N       & $2 \times 10^{-9}$ \\
\C2H3CN{}     & $3\pm0.6 \times 10^{-8}$  \\
HCN           & $1.4\pm0.3 \times 10^{-7}$ \\
H$^{13}$CN           & $1.8\pm0.4 \times 10^{-8}$  \\ 
HC$^{15}$N           & $1.1\pm0.3 \times 10^{-9}$  \\ 
\hline
\hline
Parameter        & Value \\
\hline
$n_{e0}$ & $6.4\pm1.3\times10^6$ \cmc{} \\
$\ro$  & \arcsa{0}{22}$\pm$\arcsa{0}{04} \\ 
$T_0$  & $440\pm90$ K \\ 
$p_i$  & $0.8\pm0.2$ \\ 
$p_o$  & $1.8\pm0.4$ \\ 
%$\ro$  & \arcsa{0}{2} \\ 
$n_0$  & $4.0\pm0.8\times10^8$ \cmc{} \\ 
$v_0$  &4.9$\pm0.5$ \vkm{} \\
$\kappa_\nu$  & $0.022\pm0.004$ cm$^2$ g$^{-1}$ \\
\enddata
\tablenotetext{\mbox{}}{The uncertainties are assumed to be 20\% for all parameters.}
\end{deluxetable}

\clearpage

\begin{figure} [!hbp]
\centering
\putfig{0.85}{0}{f3.ps}
\figcaption[]
{Spectra toward the dense core, averaged over a circular region with a diameter
of \arcsa{0}{5}. Note that three small frequency intervals are empty,
because of bad data and an incorrect correlator setup during the
observations.
\label{fig:spec}
}
\end{figure}

\begin{figure} [!hbp]
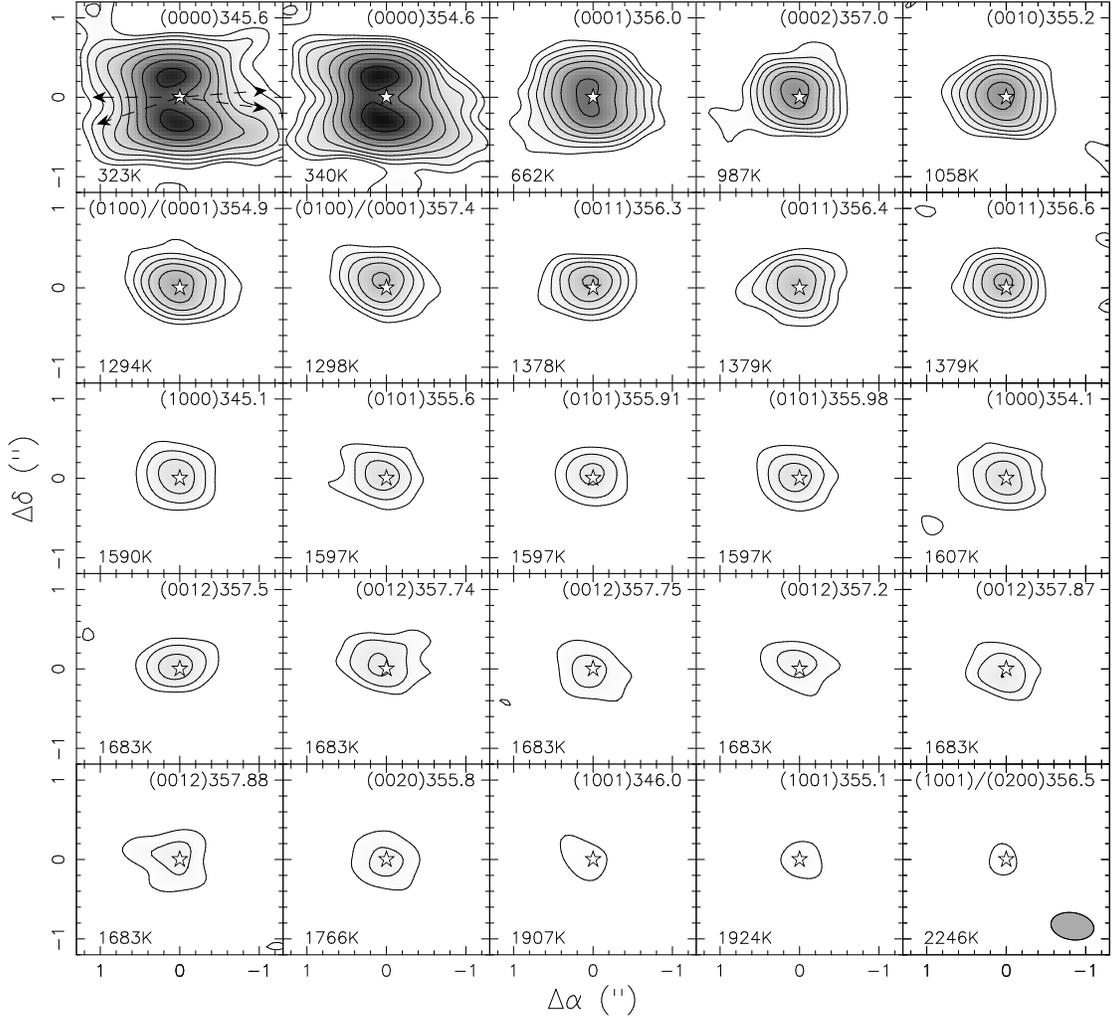

\centering
\putfig{0.8}{270}{f4.ps}
\figcaption[]
{
Integrated intensity maps of the \nHC3N{} lines toward the dense core in the order of
increasing upper energy level.  The contour levels are $A (1-r^n)/(1-r)$,
where $A=1$ \Jybk{}, $r=1.4$, and $n=1,2,3,..$.  The star marks the central star
position and the arrows indicate the outflow axes.  In each panel, the upper
right corner gives the vibrational state $(v_4 v_5 v_6 v_7)$ in parenthesis and rest frequency
in GHz, and the lower left corner gives the upper energy level in K.
The resolution is \arcsa{0}{54}$\times$\arcsa{0}{34}.
\label{fig:HC3N}
}
\end{figure}

\begin{figure} [!hbp]
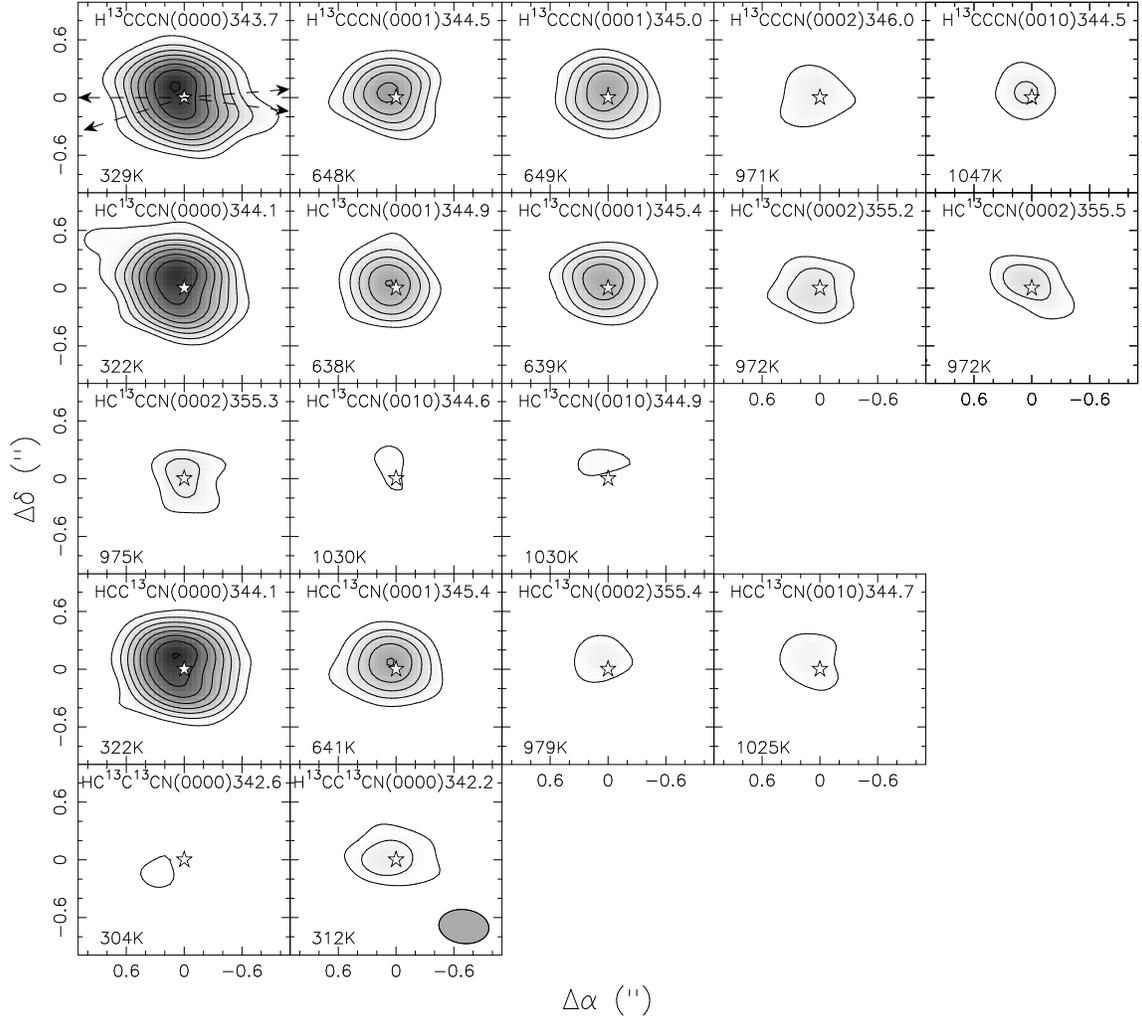

\centering
\putfig{0.8}{270}{f5.ps}
\figcaption[]
{Same as that in Fig. \ref{fig:HC3N} but for the lines of the \nHC3N{}
isotopologues.  The contour levels are $A (1-r^n)/(1-r)$,
where $A=1$ \Jybk{}, $r=1.2$, and $n=1,2,3,..$.
\label{fig:HC3Niso}
}
\end{figure}

\begin{figure} [!hbp]
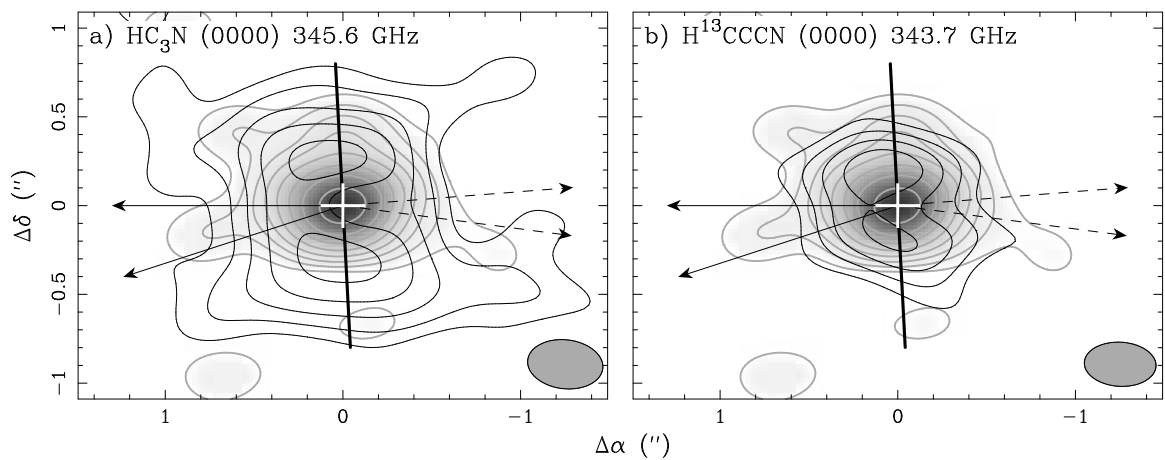

\centering
\putfig{0.65}{270}{f6.ps}
\figcaption[]
{Integrated intensity maps of the (a) \nHC3N{} (0000) and (b) \H13CCCN (0000) lines
on top of the continuum map shown in Fig.  \ref{fig:cont}b, showing the
structure of the dense core in the outer part.  The cross marks the central star
position and the arrows indicate the outflow axes.  The solid line indicates
the major axis of the dense core. The resolution is $\sim$
\arcsa{0}{42}$\times$\arcsa{0}{26}.  The contour levels are $A (1-r^n)/(1-r)$,
with $n=1,2,3,..$.  In (a) $A=2$ \Jybk{} and $r=1.4$.  In (b) $A=1.2$
\Jybk{} and $r=1.3$.  
\label{fig:HC3N_cont} 
}
\end{figure}

\begin{figure} [!hbp]
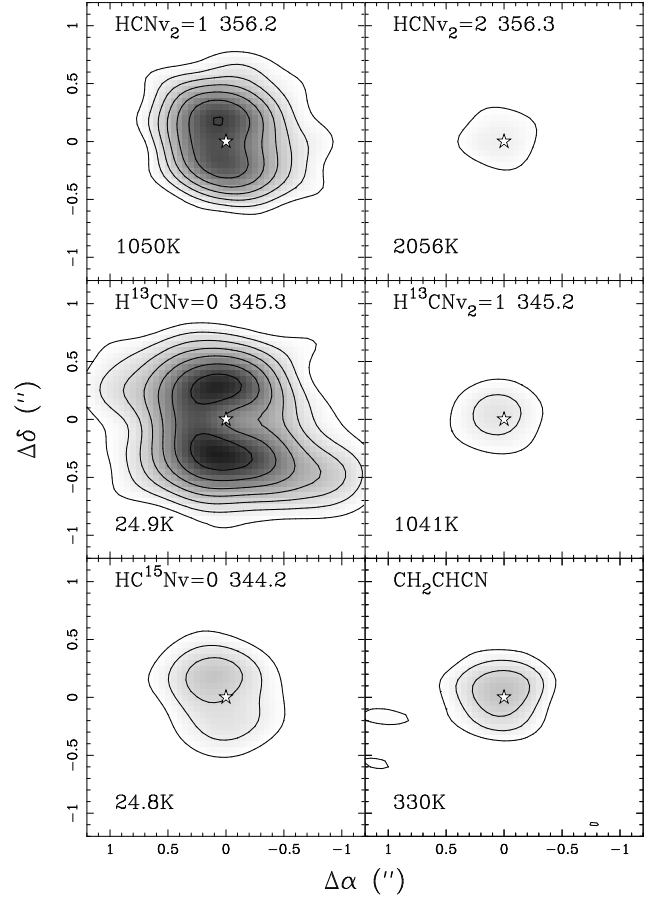

\centering
\putfig{0.7}{270}{f7.ps}
\figcaption[]
{Same as that in Fig. \ref{fig:HC3N} but for the lines of HCN and its
isotopologues that trace the dense core.  Also \C2H3CN{} map is presented
after summing over all the \C2H3CN{} lines for a higher signal to noise ratio.
The contour levels are $A (1-r^n)/(1-r)$, where $A=2$ \Jybk{},
$r=1.25$, and $n=1,2,3,..$.
\label{fig:HCN_Vinyl}
}
\end{figure}

\begin{figure} [!hbp]
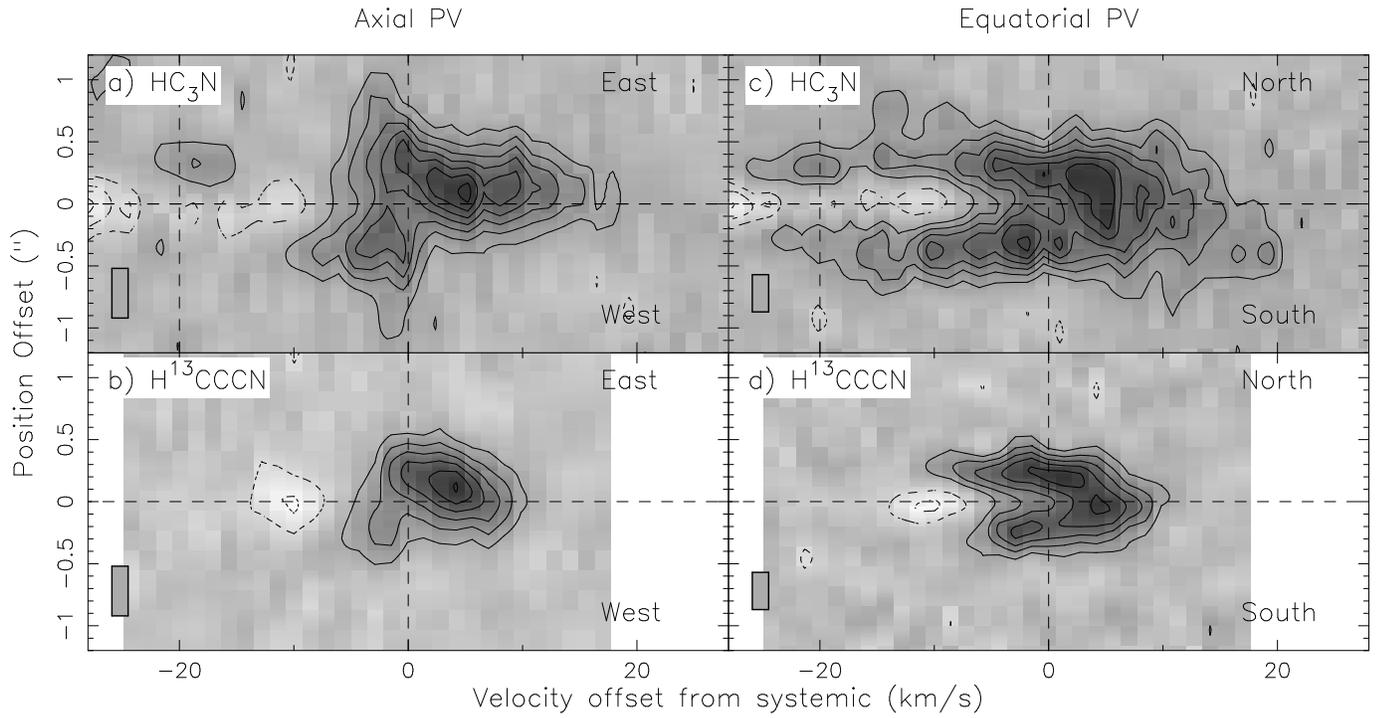

\centering
\putfig{0.75}{270}{f8.ps} %pvs_env.ps
\figcaption[]
{PV diagrams for the \nHC3N{} (0000) and \H13CCCN{} (0000) lines. (a) and (b) are the
axial PV diagrams cut perpendicular to the dense core. (c) and (d) are the
equatorial PV diagrams cut along the major axis of the dense core.
The angular and velocity resolutions are shown in the lower left corner.
The positive (negative) contours start from 16 ($-$16) K with a step of 16
($-$16) K.
\label{fig:pvs_env}
}
\end{figure}

%angular radius of the dense core

\begin{figure} [!hbp]
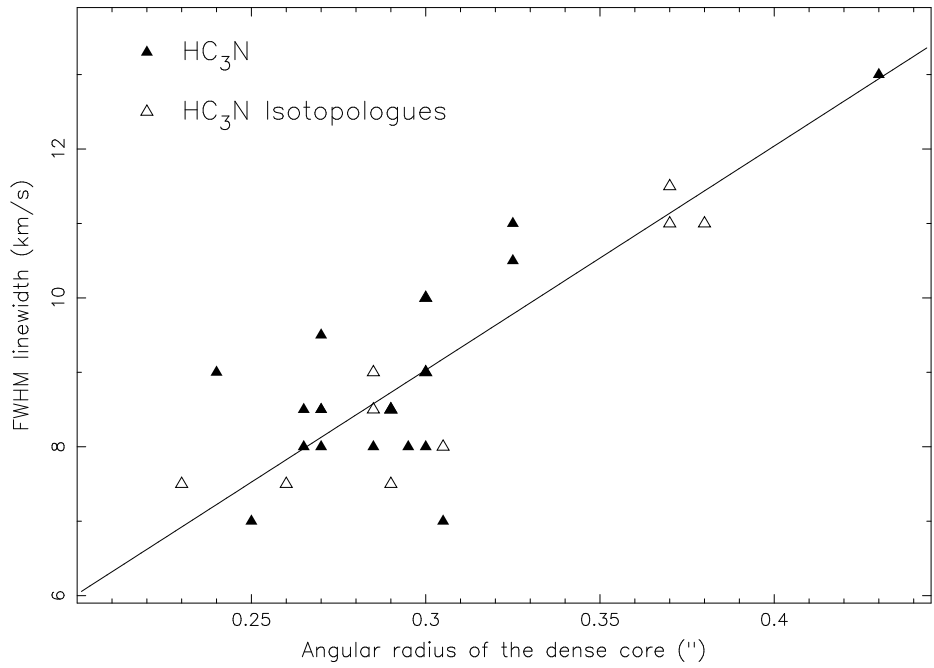

\centering
\putfig{0.5}{270}{f9.ps} % emis_size.ps
\figcaption[]
{A plot of the FWHM linewidth versus the 
half width of the emitting size (or angular radius) of the dense
core for the lines of \nHC3N{} and its singly $^{13}$C substituted
isotopologues. The solid line is a linear fit to the data.
\label{fig:emis_size}
}
\end{figure}

\begin{figure} [!hbp]
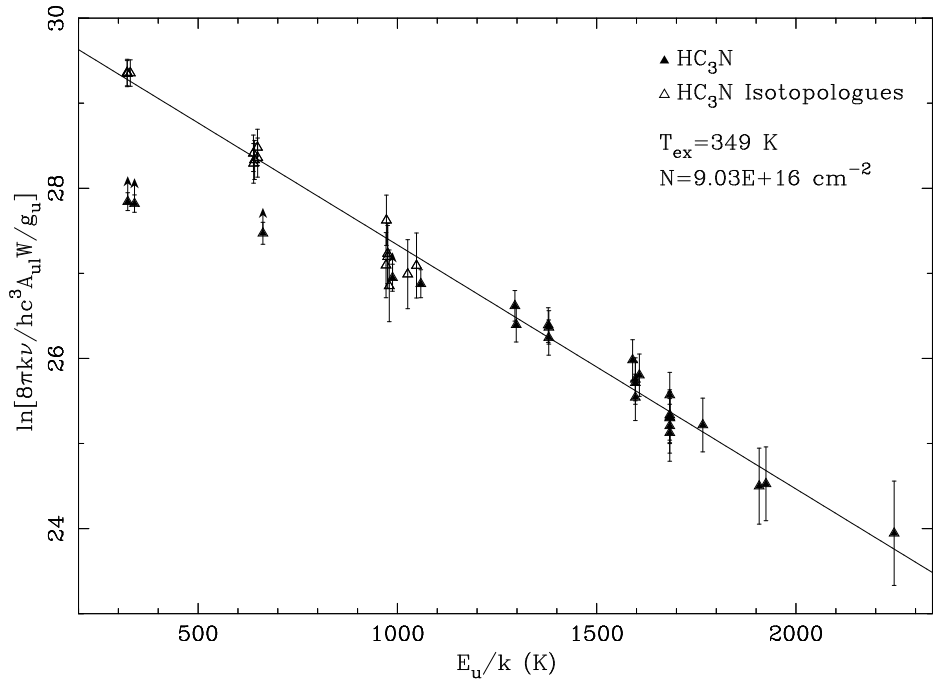

\centering
\putfig{0.5}{270}{f10.ps} % rotation_diagram.ps
%\putfig{0.5}{270}{emis_Temp.ps}
\figcaption[]
{Population diagram for \nHC3N{} and its singly $^{13}$C substituted
isotopologues. The line intensity of the isotopologues has been multiplied
by a factor of 10. The solid line is a linear fit to the data. 
The \nHC3N{} data points with an upper arrow are not optically thin and thus excluded from
the fitting.
%(Bottom) A plot of the upper energy level
%versus the angular radius of the dense core for the lines of \nHC3N{}
%and its singly $^{13}$C substituted
%isotopologues. The energy level of the isotopologues has been multiplied by
%a factor of 2.
\label{fig:rottemp}
}
\end{figure}

\begin{figure} [!hbp]
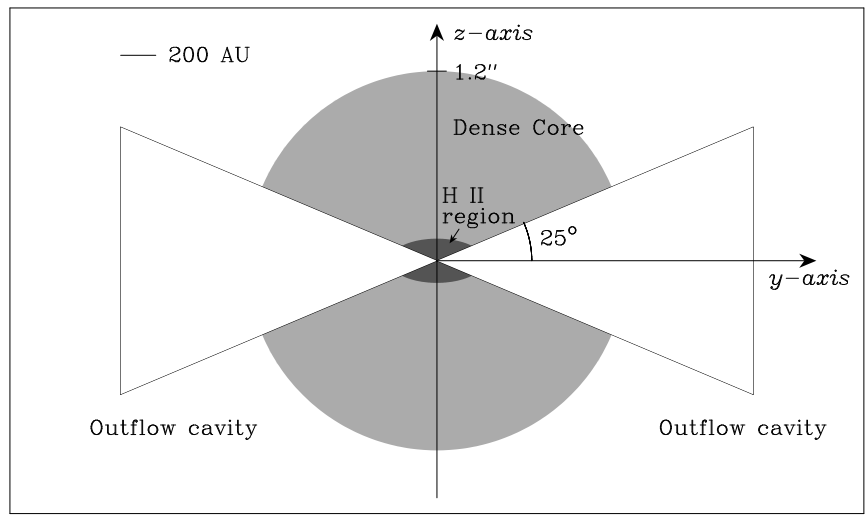

\centering
\putfig{0.5}{270}{f11.ps} %model_sketch.ps
\figcaption[]
{
A schematic diagram for our simple model in the $yz$-plane, with $y$ to the
west and $z$ to the north.  The H II region (dark gray region) is ellipsoidal with a size of
\arcsa{0}{6}$\times$\arcsa{0}{28}$\times$\arcsa{0}{28}, and thus it has a
radius of \arcsa{0}{14} in the $z$-axis and a radius of \arcsa{0}{3} in the
$y$-axis.  The dense core (light gray region) is spherical, with its outer radius set to
\arcsa{1}{2} and its inner boundary set by the outer boundary of the
ellipsoidal H II region.  There are two outflow cavities, one in the east
and one in the west, both with an half opening angle of 25\degree{}.  In our
model, the free-free emission is from the H II region, and both the dust and
molecular emissions are from the dense core.
\label{fig:modelsketch}
}
\end{figure}

\begin{figure} [!hbp]
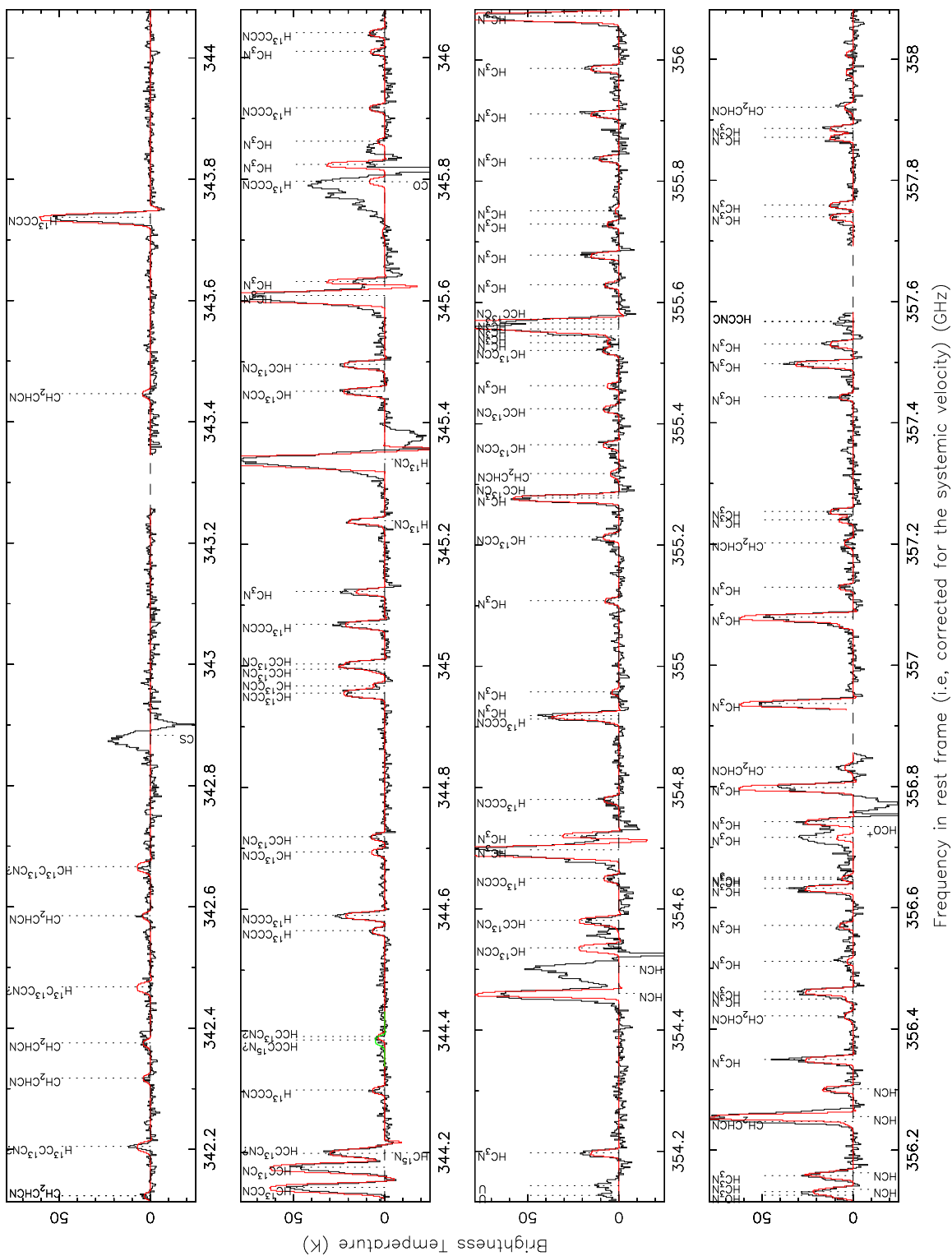

\centering
\putfig{0.85}{0}{f12.ps} % fit23.ps
\figcaption[]
{A fit to the spectra with our model. The black spectra are 
the same as those in Figure \ref{fig:spec}. Red spectra are derived from our
model.  The green spectrum shows the possible HCCC$^{15}$N (0000) line at 
344.3853481 GHz predicted from our model assuming an abundance ratio
[\nHC3N]/[HCCC$^{15}$N]=150.
\label{fig:fitspec}
}
\end{figure}

\begin{figure} [!hbp]
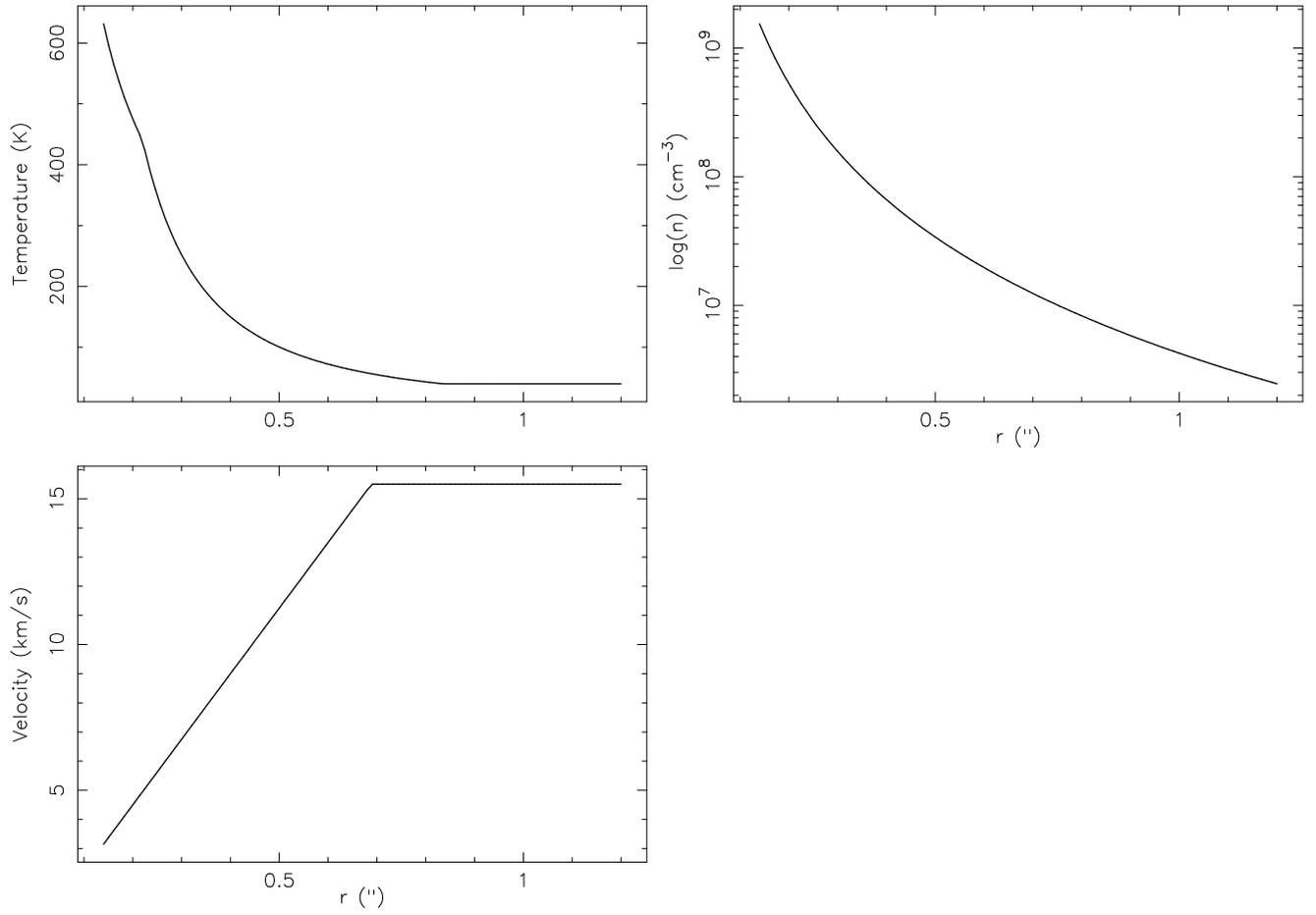

\centering
\putfig{0.7}{270}{f13.ps} % profile.ps
\figcaption[]
{The temperature, density, and velocity distributions in the dense core in our
best model.
\label{fig:modprofile}
}
\end{figure}

\begin{figure} [!hbp]
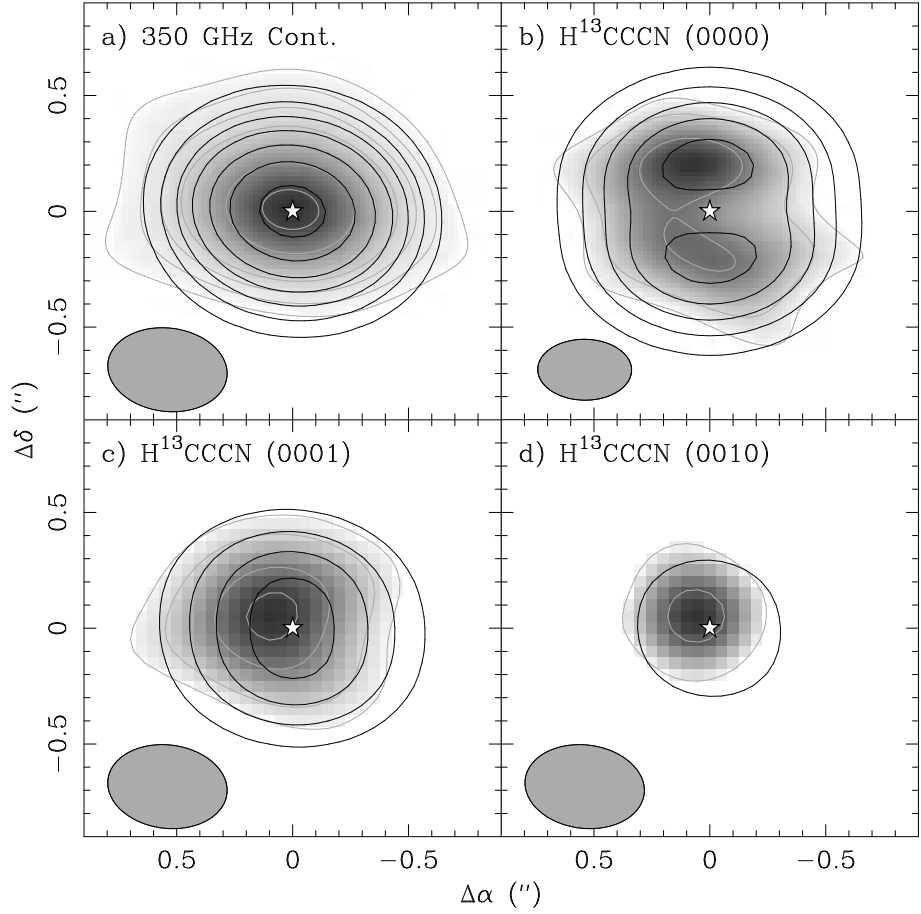

\centering
\putfig{0.7}{270}{f14.ps} % fitH13CCCN.ps
\figcaption[]
{A comparison of the emission structure in the 350 GHz continuum and three
\H13CCCN{} lines at different vibrational states.  The images with gray
contours are from the observations, and the black contours are from our
model.  Panel (a) has the same contour levels as in Fig.  \ref{fig:cont}a. 
The lowest contour is affected by the outflow and thus excluded.  Panel
(b) has the same contour levels as in Fig.  \ref{fig:HC3N_cont}b.
Panels (c) and (d) have the same contour levels as in Fig.  \ref{fig:HC3Niso}.
\label{fig:fitstr}
}
\end{figure}

\begin{figure} [!hbp]
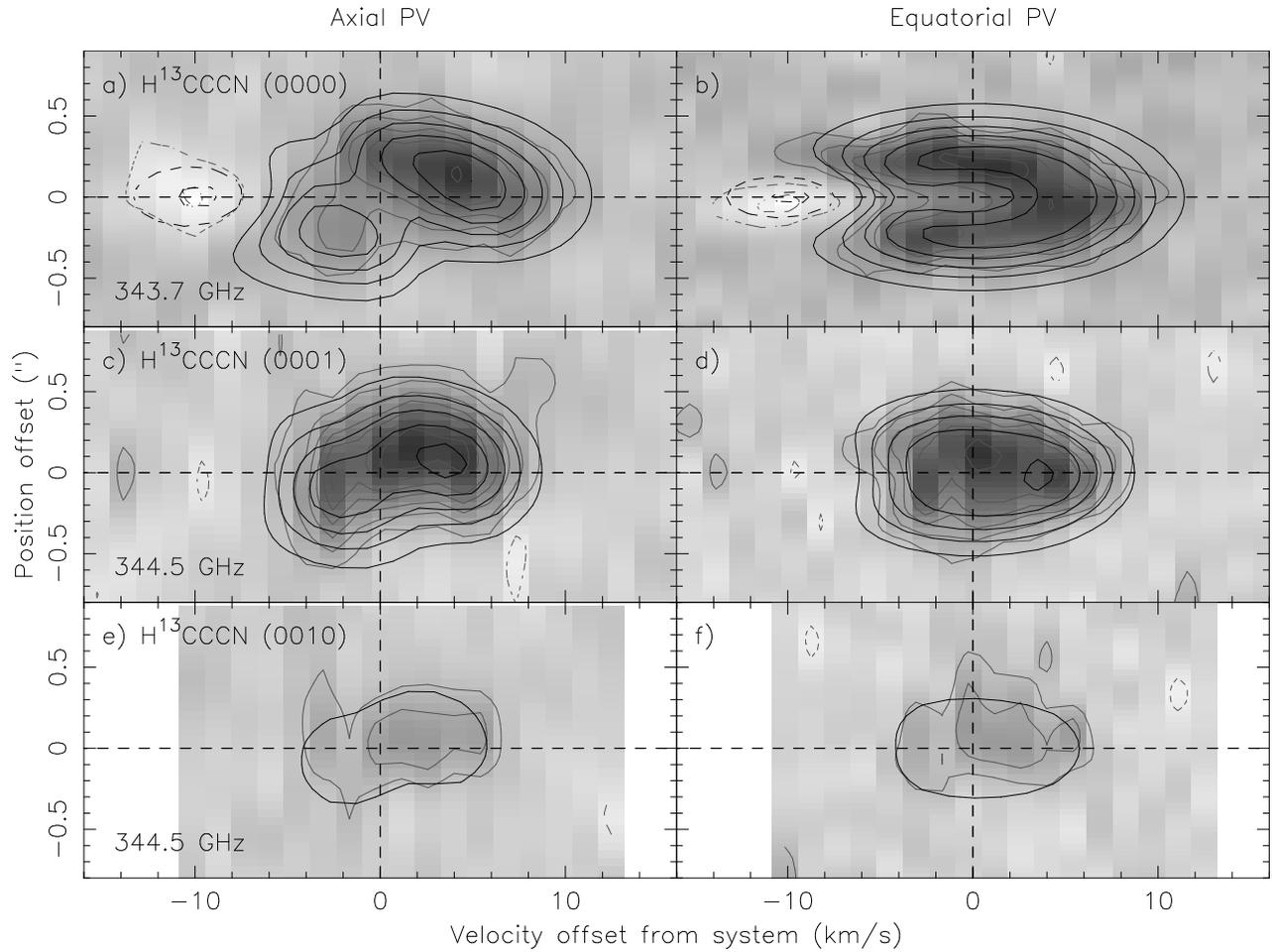

\centering
\putfig{0.7}{270}{f15.ps} % fitPVH13CCCN.ps
\figcaption[]
{A comparison of the PV structure in the three
\H13CCCN{} lines at different vibrational states.  The images with gray
contours are from the observations, and the black contours are from our
model. Panels (a) and (b) have the same contour levels as in Fig. \ref{fig:pvs_env}b.
In panels (c) to (f), the positive (negative) contours start from 6 ($-$6) K with a step of 6
($-$6) K.
\label{fig:fitPV}
}
\end{figure}

%\renewcommand{\baselinestretch}{1}
%\begin{figure}[h]
%\begin{center}
%\includegraphics[width=120mm,angle=-90]{model_sketch.eps}
%\caption[]
%{{model sketch}\label{mod_ske}}
%\end{center}
%\end{figure}
%\renewcommand{\baselinestretch}{1.4} 

\end{document}